\documentclass[a4paper,12pt]{article}
\usepackage[utf8x]{inputenc}
\usepackage{amsmath,amssymb}
\usepackage{subfigure}
\usepackage{nips12submit_e}
\usepackage{varwidth}
\usepackage{array}
\usepackage{ragged2e}
\usepackage{graphicx}
\usepackage[colorlinks=black]{hyperref}
\usepackage[colorinlistoftodos, textwidth=4cm, shadow]{todonotes}
\usepackage{amsmath}
\usepackage[displaymath, tightpage]{preview}
\newtheorem{definition}{Definition}
\newtheorem{lemma}{Lemma}

\newtheorem{example}{Example}
\newtheorem{problem}{Problem}
\usepackage{algorithm}
\title{Tripartite Graph Clustering for Dynamic Sentiment Analysis on Social Media}
\author{Linhong Zhu$^1$, Aram Galstyan$^1$, James Cheng$^2$ and Kristina Lerman$^1$\\
Information Sciences Institute, University of Southern California$^1$,\\
Dept. of Computer Science $\&$ Engineering, The Chinese University of Hong Kong$^2$\\
\{linhong, galstyan, lerman\}@isi.edu $^1$, j.cheng@acm.org$^2$}
\begin{document}

\maketitle
\tableofcontents
\section{Introduction}\label{sec:intro}



\begin{figure}[!t]
\centering
  \includegraphics[width=\columnwidth]{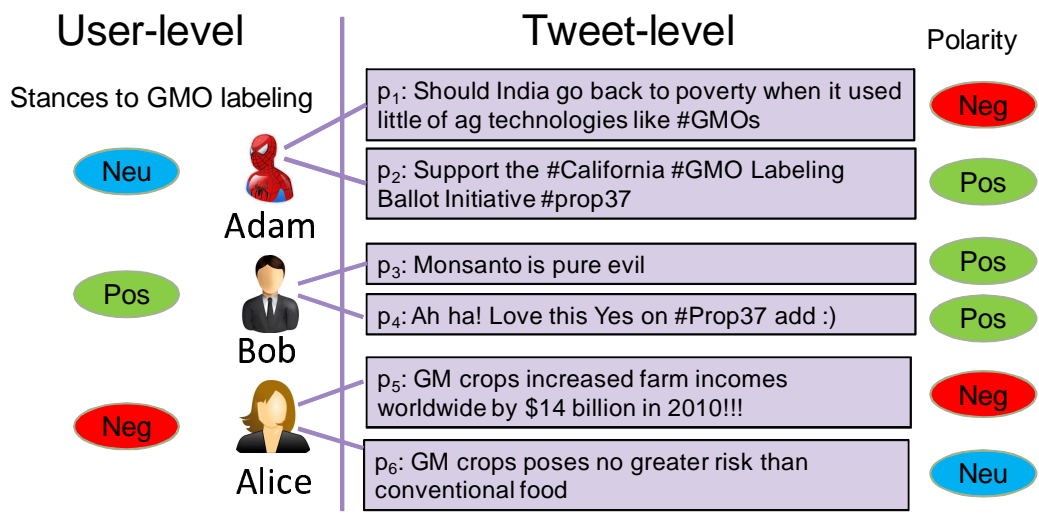}
  \caption{An example of both user-level and tweet-level sentiments toward the topic ``labeling genetically modified organism (GMO)"}\label{fig:examplesentiment}
  \vspace{-0.25cm}
\end{figure}

In the past few years, there has been a significant growth in the use of social media platforms such as Twitter. Spurred by that growth, companies, advertisers, and political campaigners are seeking ways to analyze the sentiments of users through the Twitter platform on their products, services and policies. For instance, a political group that is advocating a new policy for ``labeling genetically modified organisms (GMO)", might want to know the positions of different users toward that policy, which requires accurate estimations of users' sentiments (i.e., whether a user has expressed a positive, negative or neutral attitude towards GMO labeling), and such information can be obtained from Twitter as shown in Figure~\ref{fig:examplesentiment}.

Most prior work on Twitter sentiment analysis has focused on understanding the sentiments of individual tweets~\cite{Barbosa2010,Go2009,Davidov2010,Wangcikm2011,Speriosu:2011,WWW2013Hu}, although more recently some authors have addressed the problem of inferring user-level sentiments~\cite{Tan2011,icwsm2013}; see Section~\ref{sec:related} for more details. Smith et al.~\cite{Topics2013} and Deng et al.~\cite{confsdmDengHJLLW13} study both tweet-level and user-level sentiments, but they assume that a user's sentiment can be estimated by aggregating the sentiments of all his/her posts. Although the sentiments of users are correlated with the sentiments expressed in tweets, such simple aggregation can often produce incorrect results, in part because sentiment extracted from short texts such as tweets will generally be very noisy and error prone. For instance, in Figure~\ref{fig:examplesentiment}, Bob is positive towards GMO labeling, however, tweet $p_3$ might be classified as ``negative" due to the occurrence of word ``evil" and simple aggregation of both $p_3$ and $p_4$ would produce incorrect sentiment for Bob. In contrast, inferring the sentiment of $p_3$ jointly with the sentiment of Bob's other tweets can potentially produce a more accurate classification for tweet $p_3$ and Bob's overall sentiment. This example motivates us to jointly analyze both tweet-level sentiments and user-level sentiments, by modeling the dependencies among users, tweets and words.

%

Another important challenge is understanding and characterizing the temporal evolution of user-level sentiment. For example, user Adam in Figure~\ref{fig:examplesentiment} seems to be against GMO labeling at first, but changes his mind in support of GMO labeling, perhaps due to interactions with other users. Another real-world example is provided by the recent release of iPhone5: While many users were tweeting positive comments prior to the release, several hours of limited sales and high price issues generated a wave of negative sentiments from users. Unfortunately, current offline approaches~\cite{Barbosa2010,Go2009,Davidov2010,Wangcikm2011,Speriosu:2011,WWW2013Hu} that focus on static data, might either miss those dynamic patterns in the temporal data by simply classifying users' sentiments as neutral, or become very time-consuming when applied repeatedly to temporal snapshots (e.g., each minute/hour) of the entire collection. While recent work has addressed sentiment dynamics~\cite{Castellanos2011,NguyenWSDM2012}, those studies have mainly focused on understanding how the aggregate volume of positive/negative tweets changes with time, while discarding possible interesting dynamics on individual user level. Furthermore, due to the ``long-tail" phenomenon, changes in tweet volume might be attributed to a relatively small fraction of super-active users, so that the dynamics of the aggregate sentiment might not be very representative of a typical user.

\begin{figure}[!t]
\centering
  \includegraphics[width=0.9\columnwidth]{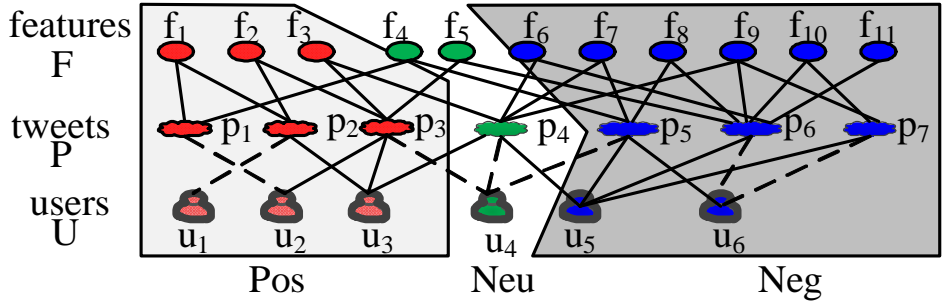}
  \caption{Co-clustering of a tripartite graph of features, users and tweets. The dashed/solid lines represent posting/re-tweeting relations between users and tweets.}\label{fig:tricoclustering}
  \vspace{-0.25cm}
\end{figure}

To address the above-mentioned challenges, we study the sentiment co-clustering problem, which aims to simultaneously cluster the sentiment of tweets and users. We propose a unified unsupervised \emph{tri-clustering framework}, to solve the sentiment-clustering problem by solving its dual problem: co-cluster a tripartite graph which represents dependencies among tweets, users and features into sentiment class. We give an example of a tripartite graph co-clustering as follows.

\begin{example}
Figure~\ref{fig:tricoclustering} shows an example of a tripartite graph, which models the correlation among features (e.g., words), tweets and users. There are three layers of nodes (i.e., features $F$, users $U$, and tweets $P$), where a feature node $f\in F$ is connected with a tweet $p\in P$ if the tweet $p$ contains the feature $f$; and a user $u\in U$ is connected with a tweet $p$ if either $u$ posts or re-tweets the tweet $p$. Therefore, if we can obtain a good clustering over the tripartite graph, for instance, we obtain three subsets \{$f_1$, $f_2$, $f_3$, $p_1$, $p_2$, $p_3$, $u_1$, $u_2$, $u_3$\}, \{$f_4$, $f_5$, $p_4$, $u_4$\} and the remaining, then the users are clustered into positive users\{$u_1$, $u_2$, $u_3$\}, neutral users\{$u_4$\} and negative users\{$u_5$, $u_6$\}, while the tweets are clustered into three subsets \{$p_1$, $p_2$, $p_3$\}, \{$p_4$\} and \{$p_5$, $p_6$, $p_7$\} simultaneously.
\end{example}

This example reveals several advantages of the proposed tri-clustering framework. First, tri-clustering framework exploits the duality between sentiment clustering and tripartite graph co-clustering to perform both user-level and tweet-level sentiment analysis. Second, since co-clustering is an unsupervised approach, neither labeled data nor high quality of labels are required, though performance can be improved by including high quality labeled data or outputs of other sentiment analysis approaches. Finally, a useful feature of co-clustering is that it can utilize the intermediate clustering results of tweets to improve the clustering results of users, and vice versa.


We present a non-negative matrix co-factor\-ization algorithm to obtain a good co-clustering of a tripartite graph. Since the cluster label corresponds to a special instance ``sentiment",  we adopt the emotion consistency regularization to make the clustering of features more relevant to the feature lexicon, and the clusters close to the sentiment classes. Motivated by prior observations that  social relation information is highly correlated with sentiment~\cite{Topics2013}, we incorporate graph regularization technique into matrix co-factorization, and use it to exploit social relationship information for  better sentiment clustering. The graph regularization aims to incur a penalty if two users are close in social relation but have different sentiments.

Finally, we extend the tripartite graph co-clustering to an online setting. We leverage previous clustering results to obtain better clustering performance for newly arrived Twitter data. Furthermore, our online framework also uses the temporal regularization over smooth evolution of both features (i.e., vocabularies) and users: (i) sentiments of vocabularies evolve smoothly over a short period; (ii) considering the entire population, the majority of users rarely change their mind within a short time. By minimizing the temporal regularization, our framework is able to achieve high accuracy in both tweet-level and user-level dynamic sentiment analysis.

We summarize the main contributions of our work as follows:
\begin{itemize}
\vspace{-0.1cm}
  \item We propose an unsupervised unified tri-clustering framework for both tweet-level and user-level sentiment analysis. We then design an analytical multiplicative algorithm to solve sentiment clustering problem with the offline tri-clustering framework.
\vspace{-0.1cm}
  \item We incorporate online setting into our framework, which allows us to study the dynamic factor of user-level sentiments, as well as the evolution of latent feature factors. Next, we also devise an online algorithm for dynamic tri-clustering problem, which is efficient in terms of both computation and storage, and effective in terms of clustering accuracy.

\vspace{-0.1cm}
  \item We conduct a set of experiments on the November 2012 California ballot Twitter data, and verify that our approach is more effective than the state-of-the-art unsupervised method for sentiment analysis, ESSA \cite{WWW2013Hu}, and is even comparable to supervised methods such as SVM~\cite{Topics2013} and Na\"{\i}ve Bayes~\cite{Go2009},  and semi-supervised methods such as Label propagation~\cite{Goldberg2006,Speriosu:2011,Tan2011} and UserReg~\cite{confsdmDengHJLLW13}.
\end{itemize}

In the rest of the paper, we first introduce the notations and formally define our problem in Section~\ref{sec:problem}. In Section~\ref{sec:offline} we propose an offline tri-clustering framework, and develop an analytical algorithm to solve the clustering problem in the offline framework. Next, in Section~\ref{sec:online} we show how to extend the offline framework into the online setting by incorporating the temporal regularization. We also describe the proposed efficient online algorithm which updates the clustering of tweets, users and features with newly arrived data and partial previous results. The experimental studies are conducted in Section~\ref{sec:expt}. We review the related works in Section~\ref{sec:related} and finally conclude our work in Section~\ref{sec:conclude}.

\section{Problem definition}\label{sec:problem}

We first introduce the notations used in this paper, which are listed in Table~\ref{tab:notations} for quick reference. A tweet is represented as a triple $p$=$<x$, $u$, $t>$ where $x$ is the feature vector representing the tweet $p$, $u$ represents a user who posts $p$ and $t$ is the associated timestamp. The label of sentiment classes is denoted as $c\in\{\texttt{pos, neg, neu}\}$. For easy representation, we also use $n$ to denote the number of tweets, $m$ to denote the number of users, $l$ to denote the number of features, and $k$ to denote the number of clusters/classes.

For each tweet $p_i$, its \emph{tweet-level sentiment} can be represented as a vector $S_{p(i)}\in R_{+}^{k}$ where $S_{p(ij)}$ denotes the probability that tweet $p_i$ is in sentiment class $j$. Similarly, for each user $u_i$, the \emph{user-level temporal sentiment} at time $t$ can also be represented as a vector $S_{u(i)}(t)\in R_{+}^{k}$, where $S_{u(ij)}(t)$ denotes the likelihood of user $u_i$'s sentiment in class $j$ at time $t$.


With the terminologies defined above, we now formally define our sentiment co-clustering problem as follows:
\begin{problem}\label{prob:task}
Given a series of temporal tweet data $\{p_1$, $p_2$, $\cdots$, $p_t\}$ that are related to the same topic, our purpose is to automatically and collectively infer the sentiments of all the observed tweets $S_p\in R_{+}^{n\times k}$, and the temporal sentiments of all the observed users $\{S_u(1)\in R_{+}^{m\times k}$, $S_u(2)$,$\cdots$, $S_u(t)\}$.
\end{problem}

\begin{table}[!t]
  \centering
  \caption{Notations and explanations}\label{tab:notations}
  \begin{tabular}{|l|l|}
    \hline
    Notations&Explanations\\
    \hline
    $n$ and $m$& number of tweets and users\\
    \hline
     $l$ and $k$ &number of features and clusters\\
     \hline
     $P$, $U$ and $F$&a set of tweets, users, and features\\
     \hline
     $X_p$/$X_u$&tweet-feature/user-feature matrix \\
     \hline
     $X_r$/$G_u$&user-tweet/user-user matrix \\
     \hline
     $S_p$/$S_f$/$S_u$&tweet/feature/user cluster matrix\\
     \hline
     $H_p$/$H_u$& $k\times k$ association matrix\\
     \hline
     $M_{(i)}$ & the $i^{th}$ row of matrix $M$\\
     \hline
      $\texttt{tr}(M)$/$||M||_F$ &trace/Frobenius norm of matrix $M$\\
     \hline
     $D_u$/$L_u$ &Diagonal/Laplacian matrix of $G_u$\\
     \hline
     $M(t)$ & matrix $M$ at timestamp $t$\\
     \hline
     $n(t)$/$m(t)$&number of tweets/users at time $t$\\
     \hline
     $X_{ud}$/$X_{un}$&user-feature matrix for disappeared \\
     /$X_{ue}$&users, new users, and evolving users\\
    \hline
  \end{tabular}
  \vspace{-0.3cm}
\end{table} 
\section{Offline framework}\label{sec:offline}

As discussed briefly in Section~\ref{sec:intro}, we formulate Problem~\ref{prob:task} as the co-clustering problem of a tripartite graph as shown in Figure~\ref{fig:tricoclustering}. To find a good clustering for a tripartite graph, Gao et al.~\cite{Gao2005} pointed out that the co-clustering over a tripartite graph $X$-$Y$-$Z$ can be divided into the clusterings over two bipartite graphs $X$-$Y$ and $Y$-$Z$. For instance, co-clustering over a tripartite graph shown in Figure~\ref{fig:tricoclustering} can be obtained by clustering over two bipartite graphs $F$-$P$ and $P$-$U$ separately. However, this formulation has the drawback that the dependence between features and users is ignored. Furthermore, the user-user social relation, which has been verified to be effective in sentiment analysis~\cite{Topics2013}, has not been taken into consideration either.

\begin{figure}[!t]
\centering
  \includegraphics[width=0.8\columnwidth]{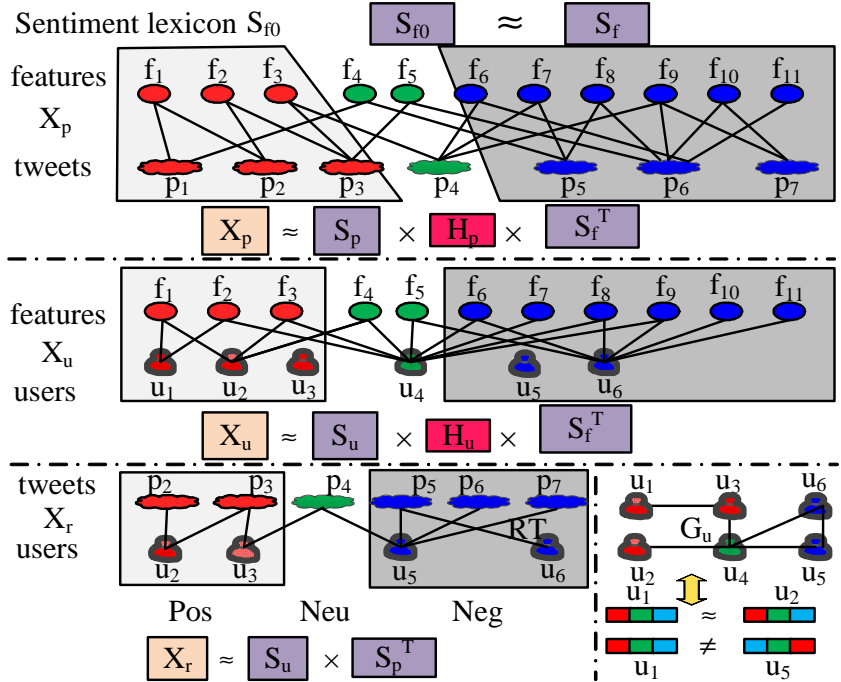}
  \caption{Offline tri-clustering framework overview (the figure is best viewed in color)}\label{fig:frame}
   \vspace{-0.3cm}
\end{figure}

To address this problem, we employ a two-level clustering for a given tripartite graph, when a tripartite graph is separated into three mutually related bipartite graphs, namely, tweet-feature bipartite graph (matrix representation $X_p\in R_{+}^{n\times l}$), user-feature bipartite graph (matrix representation $X_u\in R_{+}^{m\times l}$), and user-tweet bipartite graph (matrix representation $X_r\in R_{+}^{m\times n}$). At the inter-level, we deal with the mutual dependency among bipartite graphs, and show that the intermediate clustering results of a bipartite graph can be further applied to the clustering process of another bipartite graph. At the intra-level, we study the clustering for each single bipartite graph, and then formulate the bipartite graph clustering as a factorization problem for the non-negative matrix representation of the given bipartite graph. Finally, inter-level clustering and intra-level clustering are mutually performed.

Figure~\ref{fig:frame} illustrates the overall idea of the two-level clustering in our framework. Assume that tweets are clustered into three classes \{$p_1$, $p_2$, $p_3$\}, \{$p_4$\} and \{$p_5$, $p_6$, $p_7$\} by clustering the tweet-feature bipartite graph at the intra-level; their clustering results can then be utilized to improve the clustering of user-tweet bipartite graph at the inter-level and vise versa.

With the idea of the two-level clustering, now we propose a new Tri-clustering framework to perform both user-level and tweet-level sentiment analysis. The objective of our Tri-clustering framework is formulated as follows:
\begin{equation}
\label{equ:objective}
\small{
\begin{aligned}
\arg\min\limits_{\substack{S_f,S_u,S_p,\\H_u, H_p\geq 0}}& \{||X_p-S_pH_pS_f^{T}||_F^2 + ||X_u-S_uH_u S_f^{T}||_F^2 \\
   &+\|X_r-S_uS_p^T\|_F^2\\
     & + \alpha|| S_f- S_{f0}||_F^2   +  \beta \texttt{tr}(S_u^{T}L_{u}S_u)\} \\
 \mbox{s.t. } &  S_fS_f^T=I,S_pS_p^T=I, S_uS_u^T=I\\
\end{aligned}
}
\end{equation}
where each of the first three terms represents the intra-level bipartite graph clustering, and the aggregation of these three terms represents the inter-level bipartite graph clustering. We will discuss how simple aggregation captures the inter-level bipartite graph dependency later in Section~\ref{subsec:offlinealg}. We suggest that these three bipartite graphs are equally important and hence no parameter is introduced to control their contributions. In addition, we also incorporate sentiment lexicon information (on the top of Figure~\ref{fig:frame}) into the framework by adding regularization functions to features, and emotion correlation between users and re-tweeting users by using user-graph regularization (see right bottom of Figure~\ref{fig:frame}). Although these two pieces of information are useful, they play a minor role in sentiment clustering. Therefore, we introduce two parameters, $\alpha, \beta \in [0,1]$, to weigh the contributions of feature lexicon and user-graph regularization. In the following, we elaborate more details about each component.
\begin{itemize}
\item \textbf{co-clustering of tweets and features.}
\begin{equation}\label{equ:tweet}
\small{
\min_{S_f, H_p, S_p}||X_p-S_pH_pS_f^T||_F^2
}
\end{equation}
where $||M||_F$ denotes the Frobenius norm of a matrix $M$, $S_f\in R_{+}^{l\times k}$ denotes the feature cluster information with $S_{f(ij)}$ represents the probability that the $i$-th feature belongs to the $j$-th cluster, $H_p$ represents the association between features and tweet classes.
\item \textbf{co-clustering of users and features.}
\begin{equation}\label{equ:user}
\small{
\min_{S_f,H_u,S_u}||X_u-S_uH_uS_f^{T}||_F^2
}
\end{equation}
 where $H_u$ denotes the the association between features and user classes. This is similar to tweet clustering: we argue that users can be characterized by the word features of their tweets and word features can be clustered according to their distribution among users.
 \item \textbf{co-clustering of re-tweeting users and tweets}
\begin{equation}\label{equa:coclustering}
\small{
\min_{S_u,S_p}\|X_r-S_uS_p^T\|_F^2
}
\end{equation}
where $X_r$ is the user-retweet matrix and $X_{r(ij)}$ represents that the $i$-th user retweets the $j$-th tweet.
\item \textbf{emotion consistence between clusters and sentiment classes.}
 \begin{equation}
\small{
 \min_{S_f}||S_f - S_{f0}||_F^2
}
 \end{equation}
 where $S_{f0}$ represents the sentiment information of features (e.g. sentiment lexicon), and $S_{f0(ij)}$ is the probability that the $i$-th feature belongs to the $j$-th sentiment class. In this component, we add a regularization to make the feature representation more relevant to the task of sentiment representation, and the clusters close to the sentiment classes.
\item \textbf{emotion correlation between users and re-tweeting users.}
\begin{equation}
\small{
   \begin{aligned}
 \min_{S_u,G_u}&\frac{1}{2}\sum_{i}\sum_{j}\|S_{u(i)}-S_{u(j)}\|_2^2G_{u(i,j)}\\
  =&\texttt{tr}(S_u^{T}L_{u}S_u)
  \end{aligned}
}
\end{equation}
where  $G_u$ is a user-user graph of which each node is a user and each edge denotes the user-user re-tweeting relationship, $S_{u(i)}$ is a vector which represents the cluster association for user $i$, $L_{u}=D_u-G_u$ is the Laplacian matrix of the user-user re-tweeting graph, and \texttt{tr} represents the trace of a matrix. This equation incurs a penalty if two users are close in the user-user graph but have different sentiment labels.
\end{itemize}

\subsection{Offline Optimization Algorithm}\label{subsec:offlinealg}

In the offline framework, we develop an analytical algorithm, which belongs to the category of traditional multiplicative update algorithm~\cite{LeeNIPS2000}, to solve Eq.~(\ref{equ:objective}).
\subsubsection{Updating $\mathbf{S_f}$}
Optimizing Eq.~(\ref{equ:objective}) with respect to $S_f$ is equivalent to solving
\begin{equation*}
\small{
\begin{aligned}
\min_{S_f\geq 0}&\|X_p-S_pH_pS_f^{T}\|_F^2+ \|X_u-S_uH_u S_f^{T}\|_F^2 +\alpha\| S_f- S_{f0}\|_F^2\\
&\mbox{subject to }S_fS_f^T=I\\
\end{aligned}}
\end{equation*}
We introduce the Largrangian multiplier $\mathcal{L}$ for non-negative constraint (i.e., $S_f\geq 0$) and $\Delta$ for orthogonal constraint (i.e., $S_fS_f^T$=I) to $S_f$ in Eq.~(\ref{equ:objective}), which leads to the following Largrangian function $L(S_f)$:
\begin{equation*}
\small{
\begin{aligned}
L(S_f)&=||X_p-S_pH_pS_f^T||_F^2+||X_u-S_uH_uS_f^T||_F^2\\
&+\alpha||S_f-S_{f0}||_F^2-{\tt tr}[\mathcal{L}_{S_f}\cdot S_f^T]+\texttt{tr}[\Delta_{S_f}(S_fS_f^T-I)]\\
\end{aligned}
}
\end{equation*}
The next step is to optimize the above terms w.r.t. $S_f$. We set $\frac{\partial L(S_f)}{\partial S_f}$=0, and obtain:
\begin{equation*}
\small{
\begin{aligned}
\mathcal{L}_{S_f}=&-2X_p^T.S_pH_p+2S_fH_p^TS_p^TS_pH_p-2X_u^TS_uH_u\\
&+2S_fH_u^TS_u^TS_uH_u+2\alpha(S_f-S_{f0})+2S_f\Delta_{S_f}
\end{aligned}
}
\end{equation*}

Using the KKT condition $\mathcal{L}_{S_f}(i,j)\cdot S_f(i,j)$=0~\cite{kuhn50nonlinear}, we obtain:
\begin{equation*}
\small{
\begin{aligned}
&[-X_p^T.S_pH_p+S_fH_p^TS_p^TS_pH_p-X_u^TS_uH_u+S_fH_u^TS_u^TS_uH_u\\
&+\alpha(S_f-S_{f0})+S_f\Delta_{S_f}](i,j)S_f(i,j)=0\\
\end{aligned}
}
\end{equation*}
where $\Delta_{S_f}$ $=$ $S_f^TX_u^TS_uH_u$ $-$ $H_u^TS_u^TS_uH_u$ $+$ $Sf^TX_p^TS_pH_p$ $-$ $H_p^TS_p^T$$S_pH_p$ $-$ $\alpha S_f^T(S_f-S_{f0})$.

Following the updating rules proposed and proved in~\cite{DingKDD2006}, we have:
\begin{equation}\label{equ:updatef}
\small{
S_f\leftarrow S_f\circ\sqrt{\frac{X_u^TS_uH_u+X_p^TS_pH_p+\alpha S_{f0}+S_f\Delta_{S_f}^-}{S_fH_u^TS_u^TS_uH_u+S_fH_p^TS_p^TS_pH_p+\alpha S_f+S_f\Delta_{S_f}^+}}
}
\end{equation}
 where $\Delta_{S_f}^+=(|\Delta_{S_f}|+\Delta_{S_f})/2$ and $\Delta_{S_f}^-=(|\Delta_{S_f}|-\Delta_{S_f})/2$.

\subsubsection{Updating $\mathbf{S_p}$}
Optimizing the objective function in Eq.~(\ref{equ:objective}) with respect to $S_p$ is equivalent to solving
\[\min_{S_p\geq 0} \|X_p-S_pH_pS_f^T\|_F^2+\|X_r-S_uS_p^T\|_F^2\]
\[\mbox{subject to }S_pS_p^T=I\]

Thus, we introduce the Largrangian multiplier $\mathcal{L}$ for non-negative constraint (i.e., $S_p\geq 0$) and $\Delta$ for orthogonal constraint (i.e., $S_pS_p^T$=I) to $S_p$ in the above term, which leads to the following Largrangian function $L(S_p)$:

\begin{equation}
\begin{aligned}
L(S_p)&=\|X_p-S_pH_pS_f^T\|_F^2+\|X_r-S_uS_p^T\|_F^2\\
&-{\tt tr}[\mathcal{L}_{S_p}\cdot S_p^T]+\texttt{tr}[\Delta_{S_p}(S_pS_p^T-I)]\\
\end{aligned}
\end{equation}

The next step is to optimize the above terms w.r.t $S_p$. We set $\frac{\partial L(S_p)}{\partial S_p}$=0, and then use the Frobenius norm of a matrix $\|M\|_F^2$={\tt tr}($M^T\cdot M$), we get:

\begin{equation*}
\begin{aligned}
\mathcal{L}_{S_p}=&-2X_pS_fH_p^T+2S_pH_pS_f^TS_fH_p^T\\
&-2X_r^TS_u+2S_pS_u^TS_u+2S_p\Delta_{S_p}
\end{aligned}
\end{equation*}

Consider the KKT condition $\mathcal{L}_{S_p}(i,j)\cdot S_p(i,j)$=0~\cite{kuhn50nonlinear}, we have:
\[
[-2X_pS_fH_p^T+2S_pH_pS_f^TS_fH_p^T-2X_r^TS_u
\]

\[
+2S_pS_u^TS_u+2S_p\Delta_{S_p}](i,j)\cdot S_p(i, j)=0
\]
where $\Delta_{S_p}$=$S_p^TX_pS_fH_p^T$ - $H_pS_f^TS_fH_p^T$ + $S_p^TX_r^TS_u$ - $S_u^TS_u$.
Let $\Delta_{S_p}^+=(|\Delta_{S_p}|+\Delta_{S_p})/2$ and $\Delta_{S_p}^-=(|\Delta_{S_p}|-\Delta_{S_p})/2$, we get:
\[
[-(X_pS_fH_p^T+X_r^TS_u+S_p\Delta_{S_p}^-)
\]
\[
+(S_pH_pS_f^TS_fH_p^T+S_pS_u^TS_u+S_p\Delta_{S_p}^+)](i,j)\cdot S_p(i,j)=0
\]

Following the updating rules proposed and proofed by~\cite{DingKDD2006}, we have the updating rule of $S_p$:
\begin{equation}\label{equ:updateSp}
S_p\leftarrow S_p\circ\sqrt{\frac{X_pS_fH_p^T+X_r^TS_u+S_p\Delta_{S_p}^-}{S_pH_pS_f^TS_fH_p^T+S_pS_u^TS_u+S_p\Delta_{S_p}^+}}
\end{equation}

\subsubsection{Updating $\mathbf{S_u}$}
Optimizing the objective function in Eq.~(\ref{equ:objective}) with respect to $S_u$ is equivalent to solving
\[\min_{S_p\geq 0}\|X_u-S_uH_u S_f^{T}\|_F^2+\|X_r-S_uS_p^T\|_F^2+ \beta \texttt{tr}(S_u^{T}L_{u}S_u)\]
\[\mbox{subject to }S_uS_u^T=I\]
Thus, we introduce the Largrangian multiplier $\mathcal{L}$ for non-negative constraint (i.e., $S_u\geq 0$) and $\Delta$ for orthogonal constraint (i.e., $S_uS_u^T$=I) to $S_u$ in the above term, which leads to the following Largrangian function $L(S_u)$:

\begin{equation}
\begin{aligned}
L(S_u)&=\|X_u-S_uH_u S_f^{T}\|_F^2+\|X_r-S_uS_p^T\|_F^2+ \beta \texttt{tr}(S_u^{T}L_{u}S_u)\\
&-{\tt tr}[\mathcal{L}_{S_u}\cdot S_u^T]+\texttt{tr}[\Delta_{S_u}(S_uS_u^T-I)]\\
\end{aligned}
\end{equation}

The next step is to optimize the above terms w.r.t $S_u$. We set $\frac{\partial L(S_u)}{\partial S_u}$=0, and then use the Frobenius norm of a matrix $\|M\|_F^2$={\tt tr}($M^T\cdot M$), we get:

\[\mathcal{L}_{S_u}=-2X_uS_fH_u^T+2S_uH_uS_f^TS_fH_u^T-2X_rS_p\]
\[+2S_uS_p^TS_p+2\beta L_u S_u+2S_u\Delta_{S_u}\]

Consider the KKT condition $\mathcal{L}_{S_u}(i,j)\cdot S_u(i,j)$=0~\cite{kuhn50nonlinear}, we have:
\[[-2X_uS_fH_u^T+2S_uH_uS_f^TS_fH_u^T-2X_rS_p\]
\[+2S_uS_p^TS_p+2\beta L_uS_u+2S_u\Delta_{S_u}](i,j)\cdot S_u(i,j)=0\]
where $\Delta_{S_u}$=$S_u^TX_uS_fH_u^T$ + $S_u^TX_rS_p$ -$H_uS_f^TS_fH_u^T$ - $S_p^TS_p$ - $\beta S_u^T L_uS_u$.

Let $\Delta_{S_u}^+=(|\Delta_{S_u}|+\Delta_{S_u})/2$, $\Delta_{S_u}^-=(|\Delta_{S_u}|-\Delta_{S_u})/2$ and $L_u=D_u-G_u$, we get:

\[[-(X_uS_fH_u^T+X_rS_p+S_u\Delta_{S_u}^-+\beta G_uS_u)\]
\[+(S_uH_uS_f^TS_fH_u^T+S_uS_p^TS_p+S_u\Delta_{S_u}^++\beta D_uS_u)](i, j)\cdot S_u(i,j)=0\]

Following the updating rules proposed and proofed by~\cite{DingKDD2006}, we have the updating rule of $S_u$:
\begin{equation}\label{equ:updateSu}
S_u\leftarrow S_u\circ\sqrt{\frac{X_uS_fH_u^T+X_rS_p+\beta G_uS_u+S_u\Delta_{S_u}^-}{S_uH_uS_f^TS_fH_u^T+S_uS_p^TS_p+\beta D_uS_u)+S_u\Delta_{S_u}^+}}
\end{equation}

\subsubsection{Updating $\mathbf{H_p}$}
Optimizing the objective function in Eq.~(\ref{equ:objective}) with respect to $H_p$ is equivalent to solving
\[\min_{H_p\geq 0}\|X_p-S_pH_pS_f^T\|_F^2\]
Let $\mathcal{L}_{H_p}$ be the Lagrange multiplier for constraint $H_p\geq 0$, the Lagrange function $L(H_p)$ is defined as follows:
\[L(H_p)=\|X_p-S_pH_pS_f^T\|_F^2-tr(\mathcal{L}_{H_p}\cdot H_p^T)\]
To optimize the above term w.r.t $H_p$, we set  $\frac{\partial L(H_p)}{\partial H_p}$=0, we get:
\[
\mathcal{L}_{H_p}=-2S_p^TX_pS_f+2S_p^TS_pH_pS_f^TS_f
\]

Considering the KKT condition $\mathcal{L}_{H_p}(i,j)\cdot H_p(i,j)$=0~\cite{kuhn50nonlinear}, we have:
\[
[-2S_p^TX_pS_f+2S_p^TS_pH_pS_f^TS_f](i,j)\cdot H_p(i,j)=0
\]

Following the updating rules proposed and proofed by~\cite{DingKDD2006}, we have the updating rule of $H_p$:

\begin{equation}\label{equ:updateHp}
H_p\leftarrow H_p\circ\sqrt{\frac{S_p^TX_pS_f}{S_p^TS_pH_pS_f^TS_f}}
\end{equation}
\subsubsection{Updating $\mathbf{H_u}$}
Optimizing the objective function in Eq.~(\ref{equ:objective}) with respect to $H_u$ is equivalent to solving
\[\min_{H_u\geq 0}\|X_u-S_uH_uS_f^T\|_F^2\]
Let $\mathcal{L}_{H_u}$ be the Lagrange multiplier for constraint $H_u\geq 0$, the Lagrange function $L(H_u)$ is defined as follows:
\[L(H_u)=\|X_u-S_uH_uS_f^T\|_F^2-tr(\mathcal{L}_{H_u}\cdot H_u^T)\]
To optimize the above term w.r.t $H_u$, we set  $\frac{\partial L(H_u)}{\partial H_u}$=0, we get:
\[
\mathcal{L}_{H_u}=-2S_u^TX_uS_f+2S_u^TS_uH_uS_f^TS_f
\]
Considering the KKT condition $\mathcal{L}_{H_u}(i,j)\cdot H_u(i,j)$=0~\cite{kuhn50nonlinear}, we have:
\[
[-2S_u^TX_uS_f+2S_u^TS_uH_uS_f^TS_f](i,j)\cdot H_u(i,j)=0
\]
Following the updating rules proposed and proofed by~\cite{DingKDD2006}, we have the updating rule of $H_u$:
\begin{equation}\label{equ:updateHu}
H_u\leftarrow H_u\circ\sqrt{\frac{S_u^TX_uS_f}{S_u^TS_uH_uS_f^TS_f}}
\end{equation}

\begin{algorithm}[!t]
\small{
\caption{offline algorithm for tri-clustering}\label{alg:offline}
\begin{tabbing}
\textbf{Input}: $X_p$, $X_r$, $X_u$, $S_{f0}$, user-user retweeting graph $G_u$,\\
\hspace{1.05cm} parameters: $\alpha$ and $\beta$\\
\textbf{Output}: $S_u$, $S_p$, $S_f$\\
1: initialize $S_u$, $S_p$, $S_f$, $H_u$, $H_p\geq 0$;\\
2: \textbf{while} not converge:\\
3: \hspace{0.5cm}update $S_p$ according to Eq.~(\ref{equ:updateSp});\\
4: \hspace{0.5cm}update $H_p$ according to Eq.~(\ref{equ:updateHp});\\
5: \hspace{0.5cm}update $S_u$ and $H_u$ according to Eq.~(\ref{equ:updateSu}) and Eq.~(\ref{equ:updateHu});\\
6: \hspace{0.5cm}update $S_f$ according to Eq.~(\ref{equ:updatef});\\
7: \textbf{return} $S_u$, $S_p$, $S_f$;
\end{tabbing}
}
\end{algorithm}

Algorithm~\ref{alg:offline} gives the offline optimization algorithm. From Algorithm~\ref{alg:offline}, we can understand better why simple aggregation in Eq.~(\ref{equ:objective}) has taken advantage of the mutual relations among tweets, features and users. For example, in Line~5, updating sentiment clustering of users $S_u$ requires the intermediate results of tweet clustering $S_p$ and feature clustering $S_f$; while in Line 6 the feature-cluster matrix $S_f$ is computed from these two matrices $S_u$ and $S_p$.

\subsection{Correctness and Convergence analysis}\label{subsec:proof}
The objective function in Eq.~(\ref{equ:objective}) has lower bound zero. To proof Algorithm. 1 converges,  we need to show that  Eq.~\ref{equ:objective} is non-increasing under the updating steps in Eq.~(\ref{equ:updatef}), Eq.~(\ref{equ:updateSp}), Eq.~(\ref{equ:updateSu}), Eq.~(\ref{equ:updateHu}) and Equ.~(\ref{equ:updateHp}). In the following, we show that Eq.~(\ref{equ:objective}) is non-increasing under the updating step in Eq.~(\ref{equ:updatef}). The Eq.~(\ref{equ:objective}) is non-increasing under the remaining updating steps can be proofed similarly. We will follow the
similar procedure described in~\cite{LeeNIPS2000}. Our proof will make
use of an auxiliary function similar to that used in the
``Majorize--Minimize: (MM) algorithm~\cite{Hunter04MMatutorial}. We begin with
the definition of the auxiliary function.

\begin{definition}
$G(v, v\prime)$ is an auxiliary function for $H(v)$ if the
conditions
\[G(v, v\prime) \geq H(v), G(v, v) = H(v)\]
are satisfied.
\end{definition}

The auxiliary function is very useful because of the
following lemma.
\begin{lemma}\label{lemma:auxupdate}
If $G$ is an auxiliary function of $H$, then $H$ is
non-increasing under the update
\begin{equation}\label{equ:auxupdate}
v^{t+1} = \arg\min_{v}G(v, v^t)
\end{equation}
where $v^t$ denote the value of parameter $v$ which is obtained from last iteration.
\end{lemma}
\noindent \textbf{Proof:}
\[H(v^{t+1}) \leq G(v^{t+1}, v(t)) \leq G(v(t), v(t)) = H(v(t))\] $\Box$

Now we will show that the update step for $S_f$ in
Eq.~(\ref{equ:updatef}) is exactly the update in Eq.~(\ref{equ:auxupdate}) with a proper
auxiliary function.

We use $\mathcal{J}$ to denote the part of objective function in Eq.~(\ref{equ:objective}) that is related to $S_f$:

\[\mathcal{J}=\|X_p-S_pH_pS_f^{T}\|_F^2+ \|X_u-S_uH_u S_f^{T}\|_F^2 +\alpha\| S_f- S_{f0}\|_F^2+\texttt{tr}[\Delta_{S_f}(S_fS_f^T-I)]\]

\begin{lemma}\label{lemma:aux1}
Function
\[G(S_f, S_f^t)=\mathcal{J}(S_f^t)-2(X_p^TS_pH_p+X_u^TS_uH_u+\alpha S_{f0}+S_f^t\Delta_{S_f^t}^-)S_f^t(\log S_f-\log S_f^t)\]
\[+2(S_fH_u^TS_u^TS_uH_u+S_fH_p^TS_p^TS_pH_p+\alpha S_f^t+S_f^t\Delta_{S_f^t}^+)(\frac{S_f^2+(S_f^t)^2}{2S_f^t}-S_f^t)\]
is an auxiliary function for $\mathcal{J}$.
\end{lemma}

\noindent \textbf{Proof}: Since $G(S_f, S_f)=\mathcal{J}(S_f)$ is obvious, we only need to show that $G(S_f, S_f^t)\geq J(S_f)$. To do this, we first compare $G(S_f, S_f^t)$ with the Taylor series expansion of $\mathcal{J}$.

The Taylor series expansion of $\mathcal{J}$ gives us:
\[
\mathcal{J}(S_f)=\mathcal{J}(S_f^t)+\mathcal{J}\prime(S_f^t)(S_f-S_f^t)
\]

where \[\mathcal{J}\prime(S_f)=\frac{\partial \mathcal{J}}{\partial S_f}=-2X_p^TS_pH_p+2S_fH_p^TS_p^TS_pH_p-2X_u^TS_uH_u\]
\[+2S_fH_u^TS_u^TS_uH_u+2\alpha(S_f-S_{f0})+2S_f\Delta_{S_f}\]

Let $\Delta_{S_f}^+=(|\Delta_{S_f}|+\Delta_{S_f})/2$ and $\Delta_{S_f}^-=(|\Delta_{S_f}|-\Delta_{S_f})/2$, this is equivalent to find:
\begin{equation}\label{equ:upperbound}
\begin{aligned}
&[-2(X_p^TS_pH_p+X_u^TS_uH_u+\alpha S_{f0}+S_f^t\Delta_{S_f^t}^-)(S_f-S_f^t)\\
&+2(S_f^tH_u^TS_u^TS_uH_u+S_f^tH_p^TS_p^TS_pH_p+\alpha S_f^t+S_f^t\Delta_{S_f^t}^+)(S_f-S_f^t)]\leq\\
&[-2(X_p^TS_pH_p+X_u^TS_uH_u+\alpha S_{f0}+S_f^t\Delta_{S_f^t}^-)S_f^t(\log S_f-\log S_f^t)\\
&+2(S_f^tH_u^TS_u^TS_uH_u+S_f^tH_p^TS_p^TS_pH_p+\alpha S_f^t+S_f^t\Delta_{S_f^t}^+)(\frac{S_f^2+(S_f^t)^2}{2S_f^t}-S_f^t)]\\
\end{aligned}
\end{equation}

We further show that expression~\ref{equ:upperbound} holds due to the following two inequality expressions:

\begin{equation}\label{equ:convex}
-(S_f-S_f^t)\leq -S_f^t(\log S_f -\log S_f^t)
\end{equation}

\begin{equation}\label{equ:yang}
S_f\leq \frac{S_f^2+(S_f^t)^2}{2S_f^t}
\end{equation}

The first inequality expression Eq.~(\ref{equ:convex}) holds due to the property of convex function $-\log x$ ($x\geq 0$). A continuously differentiable function of one variable is convex on an interval if and only if the function lies above all of its tangents: $f(x) \geq f(y) + f\prime(y)(x-y)$ for all $x$ and $y$ in the interval.
Since $-\log x$ is a convex function, and $S_f\geq 0$, if we set $x=S_f$ and $y=S_f^t$, we have
\[-\log S_f \geq -\log S_f^t -\frac{S_f-S_f^t}{S_f^t}\]
which is equivalent to Eq.~(\ref{equ:convex}).

The second inequality expression Eq.~(\ref{equ:yang}) holds due to the Yang inequality expression, which gives $ab\leq \frac{a^2+b^2}{2}$.

Thus, expression~\ref{equ:upperbound} holds and $G(S_f, S_f^t)\geq J(S_f)$. $\Box$

Replacing Eq.~(\ref{equ:auxupdate}) with the auxiliary function $G(S_f, S_f^t)$, then the minimum is achieved by setting
\[\frac{\partial G(S_f, S_f^t)}{\partial S_f}=0\], which leads to

\[-2(X_u^TS_uH_u+X_p^TS_pH_p+\alpha S_{f0}+S_f^t\Delta_{S_f^t}^-)\frac{S_f^t}{S_f}+2(S_f^tH_u^TS_u^TS_uH_u+S_f^tH_p^TS_p^TS_pH_p+\alpha S_f^t+S_f^t\Delta_{S_f^t}^+)\frac{S_f}{S_f^t}=0\]
\[S_f\leftarrow S_f^t\circ\sqrt{\frac{X_u^TS_uH_u+X_p^TS_pH_p+\alpha S_{f0}+S_f^t\Delta_{S_f^t}^-}{S_f^tH_u^TS_u^TS_uH_u+S_f^tH_p^TS_p^TS_pH_p+\alpha S_f^t+S_f^t\Delta_{S_f^t}^+}}\]
This is identical to Eq.~(\ref{equ:updatef}), which proofs the correctness of the updating rule for $S_f$ in Algorithm~\ref{alg:offline}.

In addition, since $G(S_f, S_f^t)$ is an auxiliary function, based on Lemma.~\ref{lemma:auxupdate}, $\mathcal{J}$ is nonincreasing under this update rule and thus converges to local optimum.

\noindent \textbf{Complexity Analysis.} The running time for multiplying rectangular matrices (one $m\times p$-matrix with one $p\times n$-matrix), if performed na\"{\i}vely,  is $O(mnp)$. Therefore, the running time complexity of Algorithm~\ref{alg:offline} is $(rk(nl+ml+nm+m^2))$, where $r$ is the total number of iterations, and $n$, $m$, $l$ and $k$ are described in Table~\ref{tab:notations}. Note that both $k$ and $r$ are very small in our framework: $k$= 2 or 3 (positive, negative, and/or neutral), and $r$ is around 10 to 100, which is also verified in our experiments. Furthermore, due to the sparsity of matrices and those fast matrix multiplication algorithms (e.g., two $n\times n$ matrix multiplication can be computed in $O(n^{2.3727})$), the exact computational cost can be much lower than the worst case theoretical result. For the space consumption, we need to store the four input matrices with complexity $O(nl+ml+nm+m^2)$. 
\section{Online framework}\label{sec:online}

We have introduced the offline Tri-clustering framework to handle static data. We now present our online framework, where the temporal data is coming in a streaming fashion. There are two naive ways to deal with temporal data: (1)~applying the offline tri-clustering framework to the entire dataset whenever new data is added, or (2)~applying tri-clustering only to new data independently at each time interval. The first approach gives high quality clustering but is too time-consuming, while the second one is efficient but leads to poor quality.

Our online framework achieves a good tradeoff between the above two extremes and is able to study the evolution of sentiments. Before we present the online framework, we first introduce the following two observations: (1)~The frequency distribution of vocabularies changes over time; however, the sentiments of vocabularies do not change or change slowly over time. (2) Considering the entire population, the majority of users rarely change their mind within a short time.

\begin{figure}[!t]
  \includegraphics[width=\columnwidth]{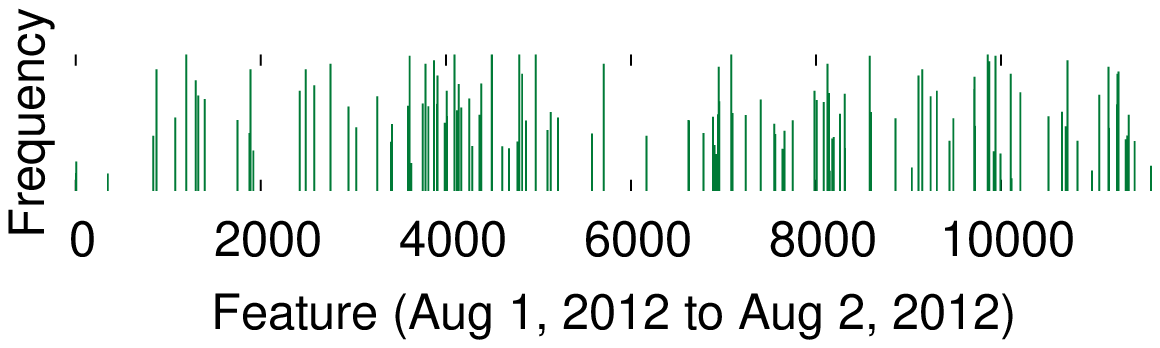}
  \includegraphics[width=\columnwidth]{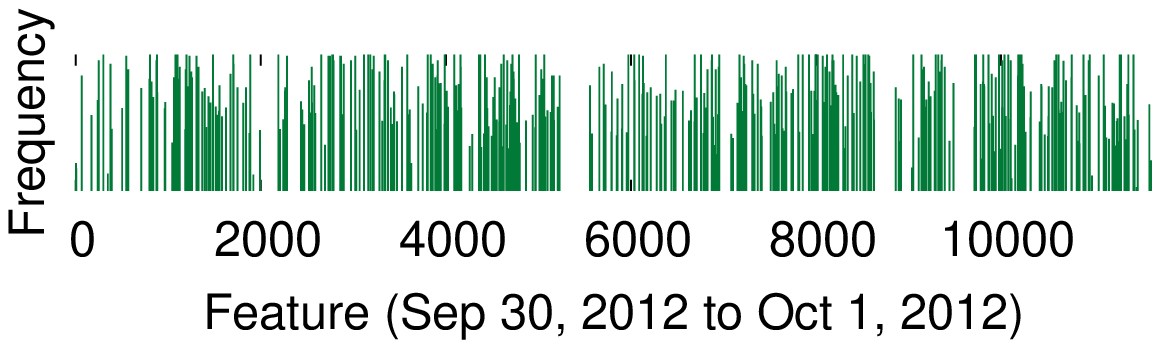}
  \caption{The evolution of features}\label{fig:featureevolution}
 \vspace{-0.3cm}
\end{figure}

To verify Observation~1, we plot the frequency of vocabularies used by the same user in two different time periods in our dataset, as shown in Figure~\ref{fig:featureevolution}. We notice that there are significant differences between the distributions of features for these two time period (the same result can be found in other randomly selected users too). Then, we show the top-8 words with the highest frequency in each pos/neg class from our dataset, as listed in Table~\ref{tab:freqword}. And we found that the set of high-frequency words tend to be popular in the entire period of our data collection. In addition, their associated sentiments also do not change over time. Thus, this observation provides us the intuition to utilize the previous sentiment clustering results of features to improve the clustering quality of tweets/users.

\begin{table}[!t]
\caption{Top-8 words with highest frequency.}\label{tab:freqword}
\centering
\begin{tabular}{|l|l|}
  \hline
Neg&\small{corn (1463),	farmer (1223), noprop37 (1211), crop (881)}\\
& \small{million (778), feed (380), India (380), seed (355)}\\
\hline
Pos&\small{yeson37 (23789), labelgmo (6485), monsanto (5809),}\\
&\small{stopmonsanto (1312), carighttoknow (1286), }\\
&\small {health (1094), safe (526), cancer (511)}\\
  \hline
\end{tabular}
\end{table}

Observation~2 has been shown in various existing works. For example, Smith et al~\cite{Topics2013} showed that the sentiments of users before election are highly correlated with the sentiments of users after election with Pearson correlated coefficient of 0.851. In addition, Deng et al.~\cite{confsdmDengHJLLW13} reported that for several topics, two posts created by the same user have similar sentiments. This is understandable since information presented in the past sets up expectations for what the user expects in the future. This observation motivates us to improve the clustering quality of users based on their previous clustering results.

Thus, intuitively, for each emerging sub-collections of tweets at time $t$, the sentiment information can be naturally obtained through factorizing new data matrix $X_p(t)$, with reference to few previous sentiment clustering results of features according to Observation~1. In other words, instead of repeatedly accessing and analyzing the past data matrices $X_p(t-1)$, $\cdots$, $X_p(1)$, we utilize intermediate sentiment clustering results of features obtained between time $[t-w,t)$, where $w$ is the time window size, to achieve good results for tweet-level sentiment analysis at time $t$.

\begin{figure}[!t]
  \includegraphics[width=\columnwidth]{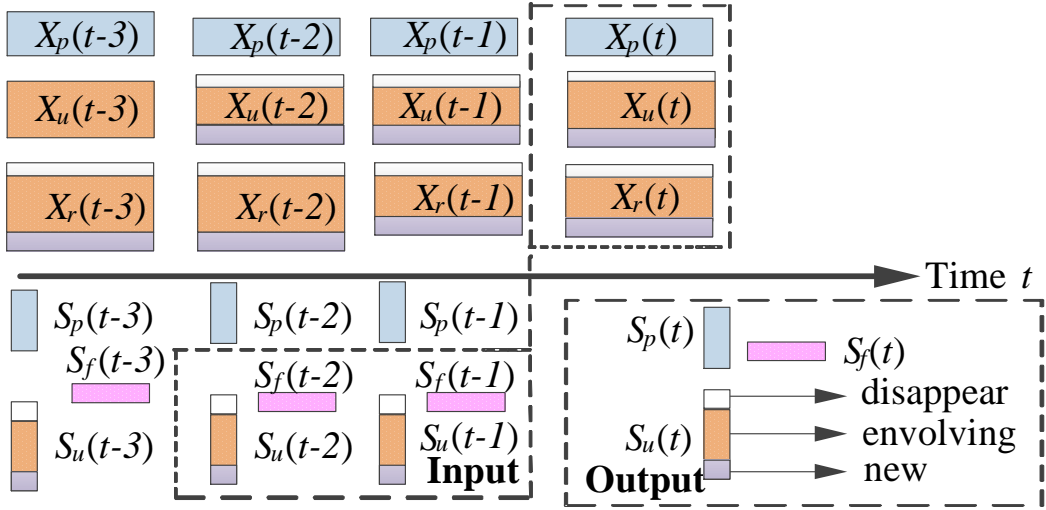}
  \caption{Online tri-clustering framework for dynamic sentiment clustering}\label{fig:onlineframe}
   \vspace{-0.3cm}
\end{figure}

Similarity, for each sub-collections of users at time $t$, we group users at time $t$ into three categories: new users, disappeared users, and evolving users. For new users, we conduct sentiment analysis based on the current user and tweet information and past sentiment clustering results of features; while for evolving users, Observation~2 suggests that previous sentiment clustering results for these users are useful for current sentiment analysis. 

Based on the above discussions, we design our online framework to make use of previous sentiment results for current sentiment analysis. Particularly, we adopt the temporal regularization technique to utilize the previous clustering results of both users and features, as suggested by the above two observations. We define the temporal regularization of a time-dependant matrix $M(t)$ to measure the smoothness of evolution as follows:
\begin{equation}\label{equ:temporal}
\small{
R(M(t))=\sum_{t}||M(t)-M_w(t)||_F^2
}
\end{equation}
where $M(t)$ denotes the matrix information at time $t$ and $M_w(t)$ denotes the past information of matrix $M$ within [$t-w$, $t$). A larger value of Eq.~(\ref{equ:temporal}) means less smoothness of evolution.

Now we address the unsolved question in Eq.~(\ref{equ:temporal}), i.e., how to utilize the previous results for current sentiment analysis. $M_w(t)$ can be simply initialized as an aggregation of all the previous results within [$t-w$, $t$). A natural modification is to give higher importance to more recent information. For example, sufficient results within [$t-w$, $t$) are aggregated over time, and an exponential decay is used to forget out-of-date results. Thus, we define $S_{fw}(t)$=$\sum_{i=1}^{w-1}\tau^iS_f(t-i)$ and $S_{uw}(t)$=$\sum_{i=1}^{w-1}\tau^iS_f(t-i)$, where $\tau\in(0,1]$ is the time-decaying factor.

The discussion above motivates the following objective function that is optimized at every time point $t$ (Figure~\ref{fig:onlineframe} depicts the overall online framework with $w$=2):
\begin{equation}
\label{equ:onlineobjective}
\small{
\begin{aligned}
\arg\min_{\substack{S_f(t),S_u(t)\\S_p(t),H_u(t), \\H_p(t)\geq 0}}& \{||X_p(t)-S_p(t)H_p(t)S_f(t)^{T}||_F^2\\
 &+ ||X_{u(e,n)}(t)-S_{u(e,n)}(t)H_u(t)S_f(t)^{T}||_F^2 \\
  &+\|X_{r(e,n)}(t)-S_{u(e,n)}(t)S_p(t)^T\|_F^2\\
  & + \alpha|| S_f(t)- S_{fw}(t)||_F^2\\
  &+\beta\mbox{tr}(S_u(t)^{T}L_{u}(t)S_u(t))\\
  &+\gamma||S_{u(d,e)}(t)-S_{uw}(t)||_F^2\}\\
 \mbox{s.t. } &  S_f(t)S_f(t)^T, S_p(t)S_p(t)^T, S_u(t)S_u(t)^T=I\\
\end{aligned}
}
\end{equation}
where $S_{u(d,e)}(t)$ denotes the sub matrix resulted by horizontally concatenating two blocks of users, $\alpha$ controls the contribution of temporal regularization for feature clustering, $\beta$ weighs the temporal graph regularization, and $\gamma$ controls the contribution of temporal regularization for user clustering.

\subsection{Online Algorithm}
We now present our online multiplicative update algorithm, which solves the objective function given by Eq.~(\ref{equ:onlineobjective}). Our algorithm is able to achieve the co-clustering for tweets, features, and users with one scan of the dataset, short response time and limited memory usage.

\vspace{0.2cm}
\noindent \textbf{$\mathbf{H_u(t)}$, $\mathbf{H_p(t)}$, and new tweets $\mathbf{S_p(t)}$.} First, let us consider the updating rules for $H_u$ and $H_p$. Since they are independent to the temporal regularization, we can naturally apply the same update rules in the offline framework to the subset data matrices at time $t$. Note that for simplicity, we just write $A(t)B(t)C(t)$ as $ABC(t)$.
\begin{equation}\label{equ:onlineupdateHu}
\small{
H_u(t)\leftarrow H_u(t)\circ\sqrt{\frac{S_{u(e,n)}^TX_{u(e,n)}S_f}{S_{u(e,n)}^TS_{u(e,n)}H_uS_f^TS_f}(t)}
}
\end{equation}
\begin{equation}\label{equ:onlineupdateHp}
\small{
H_p(t)\leftarrow H_p(t)\circ\sqrt{\frac{S_p^TX_pS_f}{S_p^TS_pH_pS_f^TS_f}(t)}
}
\end{equation}

Similarly, the tweet-level sentiments at current time, $S_p(t)$, can be updated using the following equation:
\begin{equation}\label{equ:onlineupdatep}
\small{
S_p(t)\leftarrow S_p(t)\circ\sqrt{\frac{X_pS_fH_p^T+X_{r}^TS_u+S_p\Delta_{S_p}^-}{S_pH_pS_f^TS_fH_p^T+S_pS_{u}^TS_{u}+S_p\Delta_{S_p}^+}(t)}
}
\end{equation}
where $\Delta_{S_p}$ $=$ $S_p^TX_pS_fH_p^T(t)$ $-$ $H_pS_f^TS_fH_p^T(t)$ $+$ $S_p^TX_r^TS_u(t)$ $-$ $S_u^TS_u(t)$.

\vspace{0.2cm}
\noindent \textbf{Evolving Features.} According to Observation 1, temporal regularization is used to ensure a smooth evolution from $S_{fw}(t)$ to $S_f(t)$. Therefore, for the optimization of $S_f(t)$, we first need to pay attention to the computation of $S_{fw}(t)$, which is the time-decaying aggregation of previous sentiment clustering results of features. Thus, we derive the update rule for $S_f(t)$  as follows:
\begin{equation}\label{equ:onlineupdatef}
\small{
\begin{aligned}
&S_f(t)\leftarrow S_f(t)\circ\\
&\sqrt{\frac{X_u^TS_uH_u+X_p^TS_pH_p+\alpha S_{fw}+S_f\Delta_{S_f}^-}{S_fH_u^TS_u^TS_uH_u+S_fH_p^TS_p^TS_pH_p+\alpha S_f+S_f\Delta_{S_f}^+}(t)}
\end{aligned}
}
\end{equation}
where $\Delta_{S_f}(t)$ $=$ $S_f(t)^TX_u(t)^TS_u(t)H_u(t)$ $-$ $H_u(t)S_u(t)^TS_u(t)$ $H_u(t)$ $+$ $F(t)^TX_p(t)^TS_p(t)H_p(t)$ $-$ $H_p(t)S_p(t)^T$$S_p(t)H_p(t)$ $-$ $\alpha S_f(t)^T(S_f(t)$ $-$ $S_{fw}(t))$.

\vspace{0.2cm}
\noindent \textbf{New Users.} For a new user, we do not have the smooth constraint for temporal evolution, and the sentiment information $S_u(t)$ at time $t$ can be obtained locally by factorizing the current data matrix $X_u(t)$ and $X_r(t)$. However, those new users might be connected to existing users through re-tweeting relations, and hence the optimization of $X_u(t)$ should be performed under the temporal graph regularization as well.
\begin{equation}\label{equ:onlineupdatenewu}
\small{
\begin{aligned}
&S_u(t)\leftarrow S_u(t)\circ\\
&\sqrt{\frac{X_uS_fH_u^T+X_rS_p+\beta G_uS_u+S_u\Delta_{S_{un}}^-}{S_uH_uS_f^TS_fH_u^T+S_uS_p^TS_p+\beta D_uS_u+S_u\Delta_{S_{un}}^+}(t)}
\end{aligned}
}
\end{equation}
where $\Delta_{S_{un}}(t)$ $=$ $S_u(t)^TX_u(t)S_f(t)H_u(t)^T$ $-$\\ $H_u(t)S_f(t)^TS_f(t)H_u(t)^T$ $+$ $S_u(t)^TX_r(t)S_p(t)$ $-$\\ $S_p(t)^TS_p(t)$ $-$ $\beta S_u(t)^TL_u(t)S_u(t)$.

\vspace{0.2cm}
\noindent \textbf{Evolving Users.} The computation of $S_u(t)$ for evolving users is similar to the optimization for new users except that we also have the smooth constraint for evolution (i.e., the sentiment information of users $S_u(t)$ has a steady transmission from the past $S_{uw}(t)$). In the following, we present the details about deriving updating rule for evolving users $S_u(t)$.

Optimizing the objective function in Eq.~(\ref{equ:onlineobjective}) with respect to existing users' sentiments $S_u$ is equivalent to solving (for simplicity, we write $ A(t)B(t)C(t)$ as $ABC(t))$:
\[\min_{S_u\geq 0}\|X_u(t)-S_uH_u S_f^{T}(t)\|_F^2+\|X_r(t)-S_uS_p^T(t)\|_F^2\]
\[+\beta \texttt{tr}(S_u^{T}L_{u}S_u(t))+\gamma||S_{u}(t)-S_{uw}(t)||_F^2\]
\[\mbox{subject to }S_uS_u^T(t)=I\]
Thus, we introduce the Largrangian multiplier $\mathcal{L}$ for non-negative constraint (i.e., $S_u\geq 0$) and $\Delta$ for orthogonal constraint (i.e., $S_uS_u^T(t)$=I) to $S_u$ in the above term, which leads to the following Largrangian function $L(S_u(t))$:

\begin{equation}
\begin{aligned}
L(S_u(t))&=\|X_u(t)-S_uH_u S_f^{T}(t)\|_F^2+\|X_r(t)-S_uS_p^T(t)\|_F^2+ \beta \texttt{tr}(S_u^{T}L_{u}S_u(t))\\
&\gamma||S_{u}(t)-S_{uw}(t)||_F^2-{\tt tr}[\mathcal{L}_{S_u(t)}\cdot S_u^T(t)]+\texttt{tr}[\Delta_{S_u(t)}(S_uS_u^T(t)-I)]\\
\end{aligned}
\end{equation}

The next step is to optimize the above terms w.r.t $S_u(t)$. We set $\frac{\partial L(S_u(t))}{\partial S_u(t)}$=0, and then use the Frobenius norm of a matrix $\|M\|_F^2$={\tt tr}($M^T\cdot M$), we get:

\[\mathcal{L}_{S_u(t)}=-2X_uS_fH_u^T(t)+2S_uH_uS_f^TS_fH_u^T(t)-2X_rS_p(t)\]
\[+2S_uS_p^TS_p(t)+2\beta L_u S_u(t)+2S_u\Delta_{S_u}(t)+2\gamma(S_{u}(t)-S_{uw}(t))\]

Consider the KKT condition $\mathcal{L}_{S_u(t)}(i,j)\cdot S_u(t)(i,j)$=0~\cite{kuhn50nonlinear}, we have:
\[[-2X_uS_fH_u^T(t)+2S_uH_uS_f^TS_fH_u^T(t)-2X_rS_p(t)+2S_uS_p^TS_p(t)\]
\[+2\beta L_uS_u(t)+2\gamma(S_{u}(t)-S_{uw}(t))+2S_u\Delta_{S_u}(t)](i,j)\cdot S_u(t)(i,j)=0\]
where $\Delta_{S_u(t)}$=$S_u^TX_uS_fH_u^T$(t) + $S_u^TX_rS_p$(t) -$H_uS_f^TS_fH_u^T$(t) - $S_p^TS_p$(t) - $\beta S_u^T L_uS_u$(t) -$\gamma S_u(t)^T(S_{u}(t)-S_u(t-1))(t)$

Let $\Delta_{S_u(t)}^+=(|\Delta_{S_u}|+\Delta_{S_u})/2$, $\Delta_{S_u(t)}^-=(|\Delta_{S_u}|-\Delta_{S_u})/2$ and $L_u=D_u-G_u$, we get:

\[[-(X_uS_fH_u^T(t)+X_rS_p(t)+S_u\Delta_{S_u}^-(t)+\beta G_uS_u(t)+\gamma S_{uw}(t))\]
\[+(S_uH_uS_f^TS_fH_u^T(t)+S_uS_p^TS_p(t)+S_u\Delta_{S_u}^+(t)+\beta D_uS_u(t))+\gamma S_u(t)](i, j)\cdot S_u(t)(i,j)=0\]

Following the updating rules proposed and proofed by~\cite{DingKDD2006}, we have the updating rule of $S_u(t)$:
\begin{equation}\label{equ:onlineupdateeu}
S_u(t)\leftarrow S_u(t)\circ\sqrt{\frac{X_uS_fH_u^T(t)+X_rS_p(t)+\beta G_uS_u(t)+S_u\Delta_{S_u}^-(t)+\gamma S_{uw}(t)}{S_uH_uS_f^TS_fH_u^T(t)+S_uS_p^TS_p(t)+\beta D_uS_u(t)+S_u\Delta_{S_u}^+(t)+\gamma S_u(t)}}
\end{equation}

Now we can present our on-line algorithm, as shown in Algorithm~\ref{alg:online}. The idea is similar to offline algorithm except that now we utilize the previous results for better initialization (line 1) and we have different update rules for $S_f(t)$ and $S_u(t)$ (Lines 4--8).

\begin{algorithm}[!t]
\small{
\caption{On-line algorithm for dynamic sentiment clustering\label{alg:online}}
\begin{tabbing}
\textbf{Input}: New data $X_p(t)$, $X_r(t)$, $X_u(t)$, user-user\\
\hspace{1.05cm}re-tweeting graph $G_u(t)$, old clustering matrix\\
\hspace{1.05cm}$S_{fw}(t)$ and $S_{uw}(t)$, parameters: $\alpha$, $\beta$, $\gamma$, $w$ and $\tau$\\
\textbf{Output}: $S_u(t)$, $S_p(t)$, $S_f(t)$, $H_u(t)$,  and $H_p(t)$\\
1: initialize $S_f(t)=S_{fw}(t)$, $S_{u(d,e)}(t)=S_{uw}(t)$;\\
2: randomly initialize $S_p(t), H_p(t), H_u(t)\geq 0$;\\
3: \textbf{while} not converge:\\
4: \hspace{0.5cm}update $S_f(t)$ according to Eq.~(\ref{equ:onlineupdatef});\\
5: \hspace{0.5cm}update $S_p(t)$, $H_p(t)$ according to Eq.~(\ref{equ:onlineupdatep}) and Eq.~(\ref{equ:onlineupdateHp});\\
6: \hspace{0.5cm}update $H_u(t)$ according to Eq.~(\ref{equ:onlineupdateHu});\\
7: \hspace{0.5cm}for new user: update according to Eq.~(\ref{equ:onlineupdatenewu});\\
8: \hspace{0.5cm}for evolving user: update according to Eq.~(\ref{equ:onlineupdateeu});\\
9: \textbf{return} $S_u(t)$, $S_p(t)$, $S_f(t)$;
\end{tabbing}
}
\end{algorithm}

\subsection{Correctness and Convergence analysis}
Since the online updating rules for $H_p(t)$, $H_u(t)$, $S_p(t)$ and $S_f(t)$ are similar to offline updating rules, thus we can use the convergence proof of Equ.~(\ref{equ:objective}) in Section~\ref{subsec:proof} to show that Eq.~(\ref{equ:onlineobjective}) is nonincreasing under the update steps for those matrices in Algorithm~\ref{alg:online}. Now we further show that Eq.~(\ref{equ:onlineobjective}) is nonincreasing under the update step for evolving users $S_u(t)$. In the following, we simply write $S_u(t)$ as $S_u$ to avoid the ambiguity with $S_u^t$ (which denotes the value of $S_u$ from last iteration).

Specifically, we use $\mathcal{J}$ to denote the part of objective function in Eq.~\ref{equ:onlineobjective} that is related to $S_u$:
\[\mathcal{J}(S_u)= \|X_u-S_uH_u S_f^{T}\|_F^2 + \|X_r-S_uS_p^T\|_F^2 + \beta \texttt{tr}(S_u^{T}L_{u}S_u)\]
\[+\gamma||S_{u}-S_{uw}||_F^2 + \texttt{tr}[\Delta_{S_u}(S_uS_u^T-I)]\]

\begin{lemma}
Function
\[G(S_u, S_u^t)=\mathcal{J}(S_u^t) - 2(X_uS_fH_u^T+X_rS_p+\beta G_uS_u^t+S_u^t\Delta_{S_u^t}^-+\gamma S_{uw}^t)S_u^t(\log S_u - \log S_u^t)\]
\[+2(S_u^tH_uS_f^TS_fH_u^T+S_u^tS_p^TS_p+\beta D_uS_u^t+S_u^t\Delta_{S_u^t}^++\gamma S_u^t)(\frac{S_u^2+(S_u^t)^2}{2S_u^t}-S_u^t)\]
is an auxiliary function for the
function $\mathcal{J}(S_u)$.
\end{lemma}

\noindent \textbf{Proof}: Similar to the proof to Lemma~\ref{lemma:aux1}, $G(S_u, S_u^t)\geq J(S_u)$ holds by using the Taylor expansion of $\mathcal{J}(S_u)$, the convex function of $-\log S_u$, and the Yang inequality $S_u\leq \frac{S_u^2+(S_u^t)^2}{2S_u^t}$. Since $G(S_u, S_u)=\mathcal{J}(S_u)$ is obvious, we conclude that $G(S_u, S_u^t)$ is an auxiliary function for the
function $\mathcal{J}(S_u)$.

Replacing Eq.~(\ref{equ:auxupdate}) with the auxiliary function $G(S_u, S_u^t)$, then the minimum is achieved by setting
\[\frac{\partial G(S_u, S_u^t)}{\partial S_u}=0\], which leads to

\[-2(X_uS_fH_u^T+X_rS_p+\beta G_uS_u^t+S_u^t\Delta_{S_u^t}^-+\gamma S_{uw}^t)\frac{S_u^t}{S_u}+2(S_u^tH_uS_f^TS_fH_u^T+S_u^tS_p^TS_p+\beta D_uS_u^t+S_u^t\Delta_{S_u^t}^++\gamma S_u^t)\frac{S_u}{S_u^t}=0\]
\[S_u\leftarrow S_u^t\circ\sqrt{\frac{X_uS_fH_u^T+X_rS_p+\beta G_uS_u^t+S_u^t\Delta_{S_u^t}^-+\gamma S_{uw}^t}{S_u^tH_uS_f^TS_fH_u^T+S_u^tS_p^TS_p+\beta D_uS_u^t+S_u^t\Delta_{S_u^t}^++\gamma S_u^t)}}\]
This is identical to Eq.~(\ref{equ:onlineupdateeu}), which proofs the correctness of the updating rule for $S_u$ in Algorithm~\ref{alg:online}.

In addition, since $G(S_u, S_u^t)$ is an auxiliary function, $\mathcal{J}$ is nonincreasing under this update rule and thus converges to local optimum.

\vspace{0.2cm}
\noindent \textbf{Complexity Analysis.} The computational complexity of Algorithm~\ref{alg:online} is $O(rk(n(t)l+m(t)l+n(t)m(t)+m(t)^2))$ and the space complexity is $O(n(t)l+m(t)l+n(t)m(t)+m(t)^2+lk+m(t-1)k)$, where $m(t)$ and $n(t)$ denote the number of new users and tweets at time $t$, respectively. We conclude that Algorithm~\ref{alg:online} is efficient and requires little memory due to the small size of $n(t)$ and $m(t)$ on average, which is also verified in Figures~\ref{subfig:onlinetime30} and~\ref{subfig:onlinetime37} in our experiments.

\section{Experiments}\label{sec:expt}

We use real Twitter dataset about ``California ballot initiatives" collected between August 2012 and December 2012~\cite{Topics2013}. Specifically, we choose two popular ballot initiatives, Propositions 30 (Temporary Taxes to Fund Education) and 37 (Genetically Engineered Foods, Labeling), as the target topics, and test our proposed approaches over labeled data which is related to these two propositions. The statistics of labeled tweets and users are reported in Table~\ref{tab:labeldata} (note that not every user has label information). In addition, we use the automatically built sentiment lexicon ``Yes" word lists and ``No" word lists~\cite{Topics2013} to initialize the feature sentiment class matrix $S_{f0}$.
\begin{table}[!t]
\caption{Statistics of tweets and users}\label{tab:labeldata}
\begin{center}
\begin{tabular}{|c|c|c|c|c|c|c|}
  \hline
\small{Prop}&\multicolumn{2}{c|}{Tweet} &\multicolumn{4}{c|}{User}\\
\cline{2-7}
&\small{$\texttt{Pos}$}&\small{$\texttt{Neg}$}&\small{$\texttt{Pos}$}&\small{$\texttt{Neg}$}&\small{$\texttt{Neu}$}&\small{$\texttt{unlabeled}$}\\
\hline
30&8777&5014&146&100&98&493\\
\hline
37&34789&2587&294&61&8&1564\\
\hline
\end{tabular}
\end{center}
\vspace{-0.3cm}
\end{table}

%

We quantitatively evaluate the performance of tri-clustering using the Normalized Mutual Information (NMI) and clustering accuracy to compare how well the discovered clusters reproduce the sentiment classes present in the data.

\vspace{0.2cm}
\noindent \textbf{Clustering Accuracy.} Given an outputted cluster $o \in C$ and with reference to a ground truth class $g\in G$, assume that we assign the outputted cluster with ground truth label using the majority vote, then the clustering accuracy of the outputted clustering $C$ on the ground truth clustering $G$ evaluates the percentage of data with correct assignments.
\begin{equation*}\label{equ:accuracy}
\small{
A(C,G)=\frac{1}{n}\sum_{o\in C}\max_{g\in G}|o\cap g|
}
\end{equation*}
where $n$ is the number of data samples.

\vspace{0.2cm}
\noindent \textbf{Normalized Mutual Information (NMI).} Given the outputted clustering $C$ and ground truth clustering $G$, the NMI is defined as:
\begin{equation*}\label{equ:NMI}
\small{
NMI(C,G)=\frac{2\times I(C;G)}{H(C)+H(G)}
}
\end{equation*}
where $H(C)$ and $H(G)$ denotes the entropy, and $I(C;G)$ is the mutual information between $C$ and $G$, which is defined as
\begin{equation*}
\small{
I(C;G)=\frac{\sum_i\sum_j p(o_i\cap g_j)\log\frac{p(o_i\cap g_j)}{p(o_i)p(g_j)}}{\sum_i\sum_j \frac{|o_i\cap g_j|}{n}\log\frac{n|o_i\cap g_j|}{|o_i||g_j|}}
}
\end{equation*}

In our experiments, for NMI and clustering accuracy, we use the benchmark implementation from the work~\cite{CaiPAMI2011}.

\vspace{0.2cm}
\noindent\textbf{Existing Methods for Comparison.} For tweet-level performance, we compare our offline tri-clustering approach with the state-of-the-art unsupervised method, ESSA~\cite{WWW2013Hu}, supervised methods, SVM~\cite{Topics2013} and Na\"{\i}ve Bayes (NB)~\cite{Go2009}, and semi-supervised methods, label propagation (LP)~\cite{Goldberg2006,Speriosu:2011} with 5\% labels (LP-5) and 10\% labels (LP-10) and UserReg~\cite{confsdmDengHJLLW13} with 10\% labels. ESSA has been shown in~\cite{WWW2013Hu} to outperform a set of existing unsupervised approaches such as lexicon-based approach MPQA~\cite{WilsonHLT2005} and document-clustering approach ONMTF~\cite{DingKDD2006}. LP is a popular semi-supervised method, while UserReg is a very recent method and has been shown in~\cite{confsdmDengHJLLW13} to outperform a set of existing semi-supervised approaches.

For user-level performance, we compare our approach with the recent unsupervised graph clustering method BACG~\cite{BACG12} which utilizes both structure (e.g., user-user graph) and content (e.g., feature representation for users) information, supervised methods SVM and Na\"{\i}ve Bayes, and semi-supervised methods LP~\cite{Tan2011} with 5\% labels (LP-5) and 10\% labels (LP-10) and UserReg with 10\% labels. For BACG and LP, we built a user-user retweeting graph, where each node represents a user, and each edge denotes the user-user retweeting relation. We then apply BACG to this graph to obtain the clustering of users, and apply standard LP algorithm on this graph with partial labeled nodes to classify the unlabeled nodes~\cite{Tan2011}. For UserReg, according to their paper, the sentiments of users are estimated by aggregating sentiments of related tweets~\cite{confsdmDengHJLLW13}.

For online framework, we compare our algorithm with two baseline algorithms, mini-batch and full-batch. The mini-batch algorithm, which applies our offline tri-clustering algorithm to each snapshot of new data matrices, can be considered as a simulation for the extreme design of online algorithm with high scalability but may sacrifice the quality. On the other hand, the full-batch algorithm, which applies the offline tri-clustering algorithm to the entire dataset whenever new data arrives at each timestamp, simulates another extreme with high quality but very time-consuming.

\subsection{Offline Performance Evaluation}
\begin{figure}[!t]
\centering
\subfigure[Clustering accuracy]{\includegraphics[width=0.5\columnwidth]{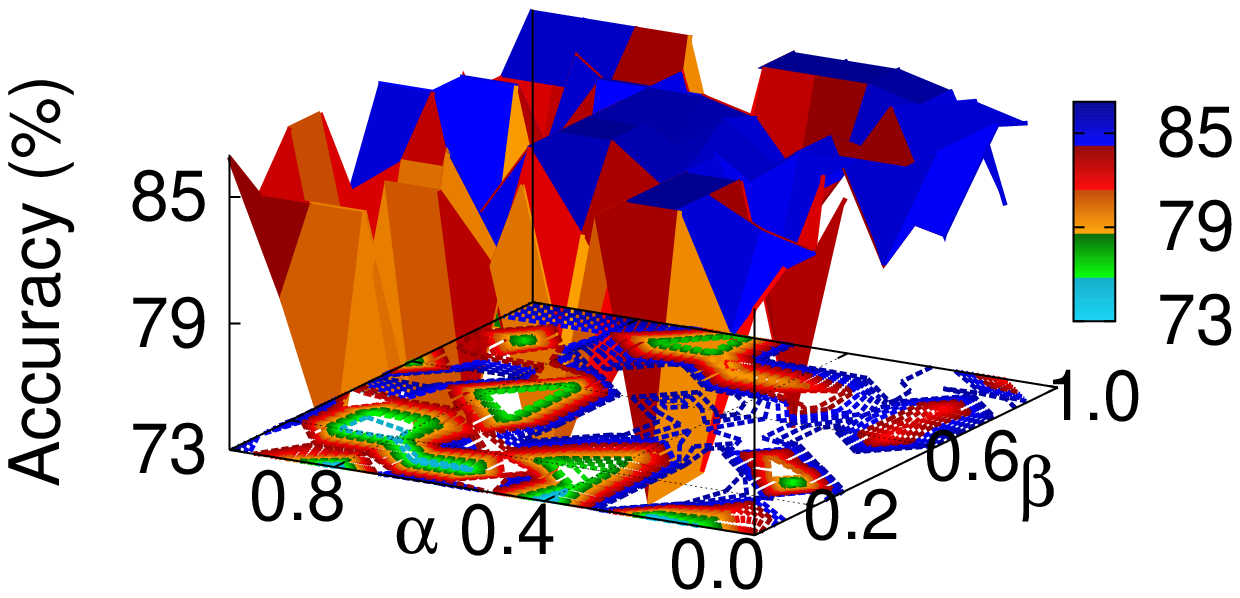}}
\hspace{-0.3cm}
\subfigure[NMI]{\includegraphics[width=0.5\columnwidth]{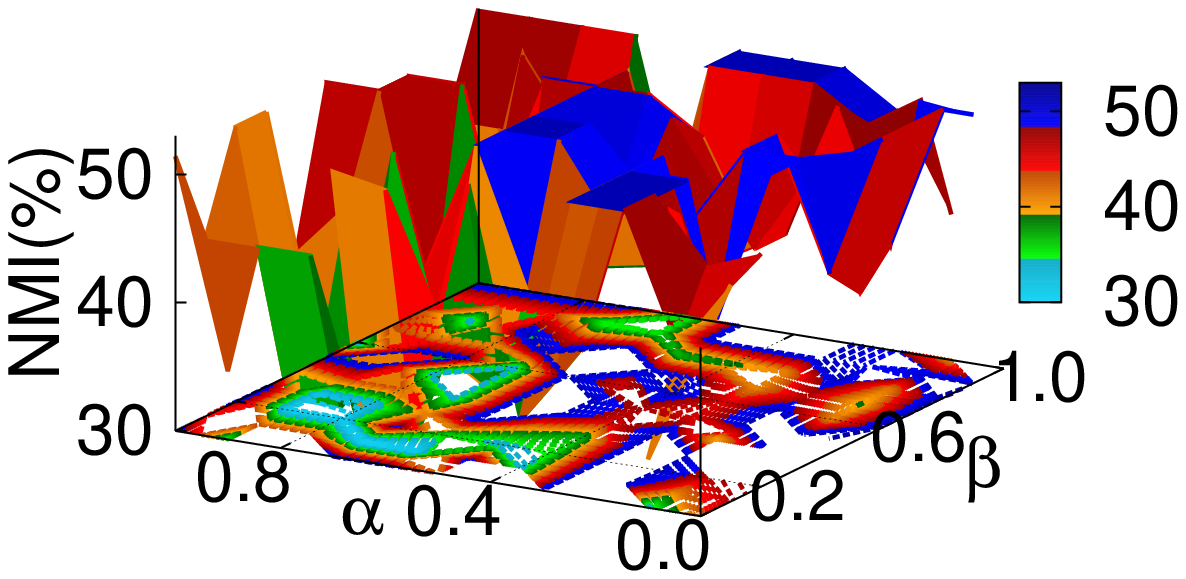}}
  \caption{User-level quality comparison when varying $\alpha$ and $\beta$ on Proposition 30 data (the figure is best viewed in color)}\label{fig:user-para}
\end{figure}
\begin{figure}[!t]
\centering
\subfigure[Clustering accuracy]{\includegraphics[width=0.495\columnwidth]{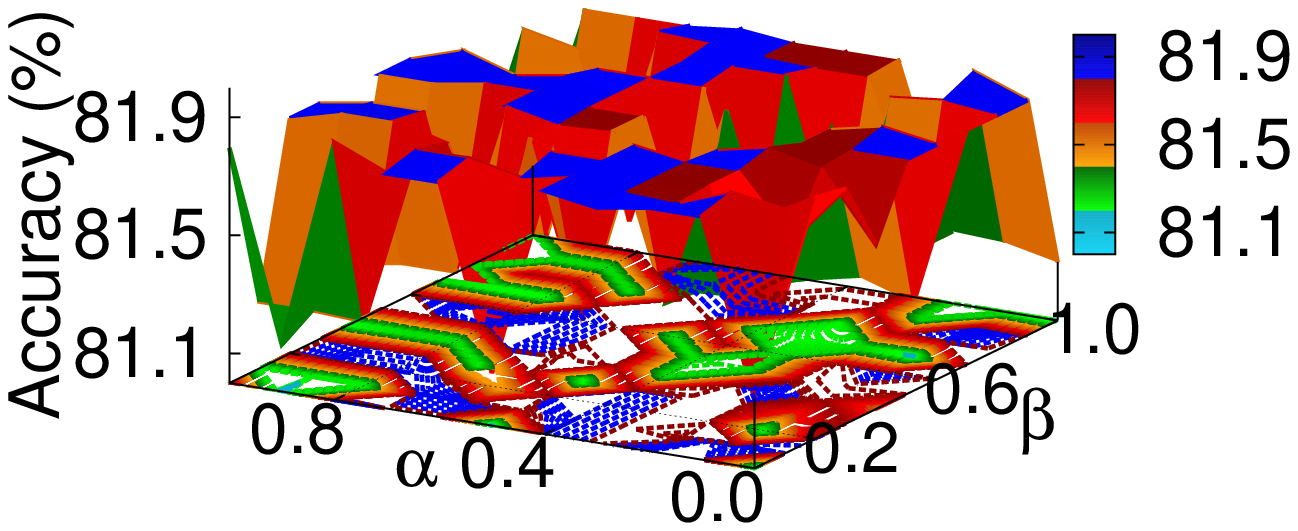}}
\hspace{-0.2cm}
\subfigure[NMI]{\includegraphics[width=0.495\columnwidth]{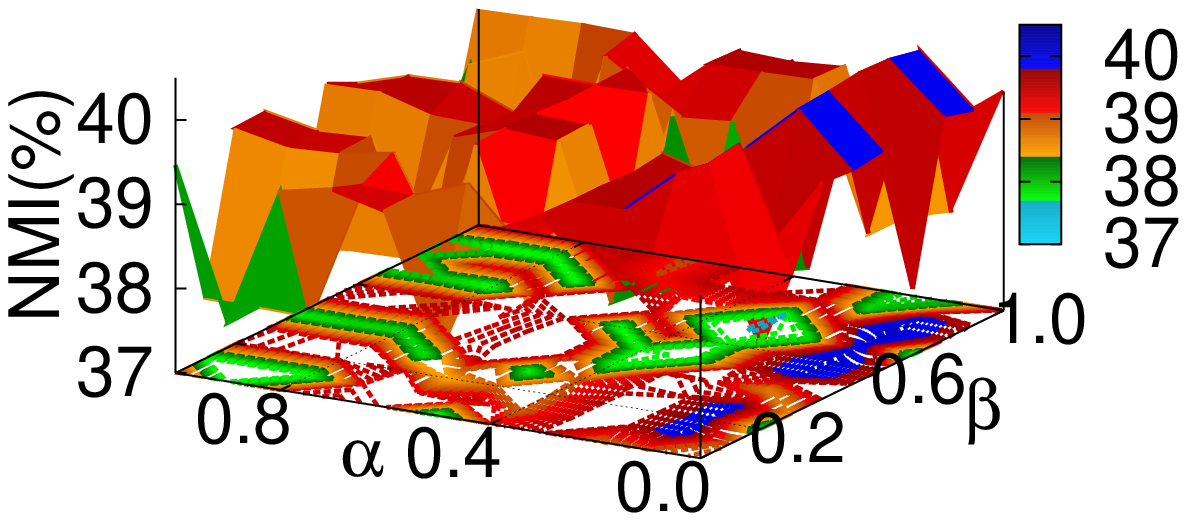}}
  \caption{Tweet-level quality comparison when varying $\alpha$ and $\beta$ on Proposition 30 data (the figure is best viewed in color)}\label{fig:tweet-para}
 \vspace{-0.3cm}
\end{figure}

\begin{figure*}[!t]
\centering
\subfigure[Average loss for Eq.~(\ref{equ:tweet})]{\includegraphics[width=0.32\columnwidth]{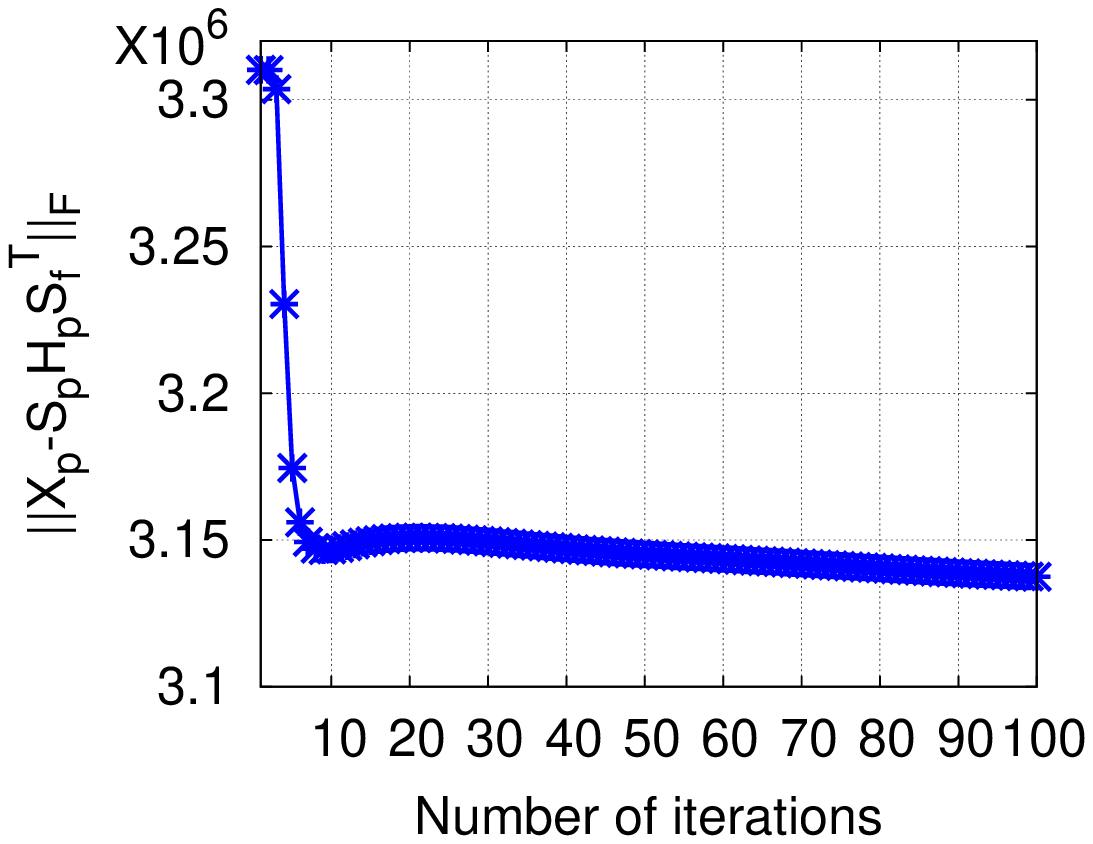}}
\hspace{-0.1mm}
\subfigure[Average loss for Eq.~(\ref{equ:user})]{\includegraphics[width=0.32\columnwidth]{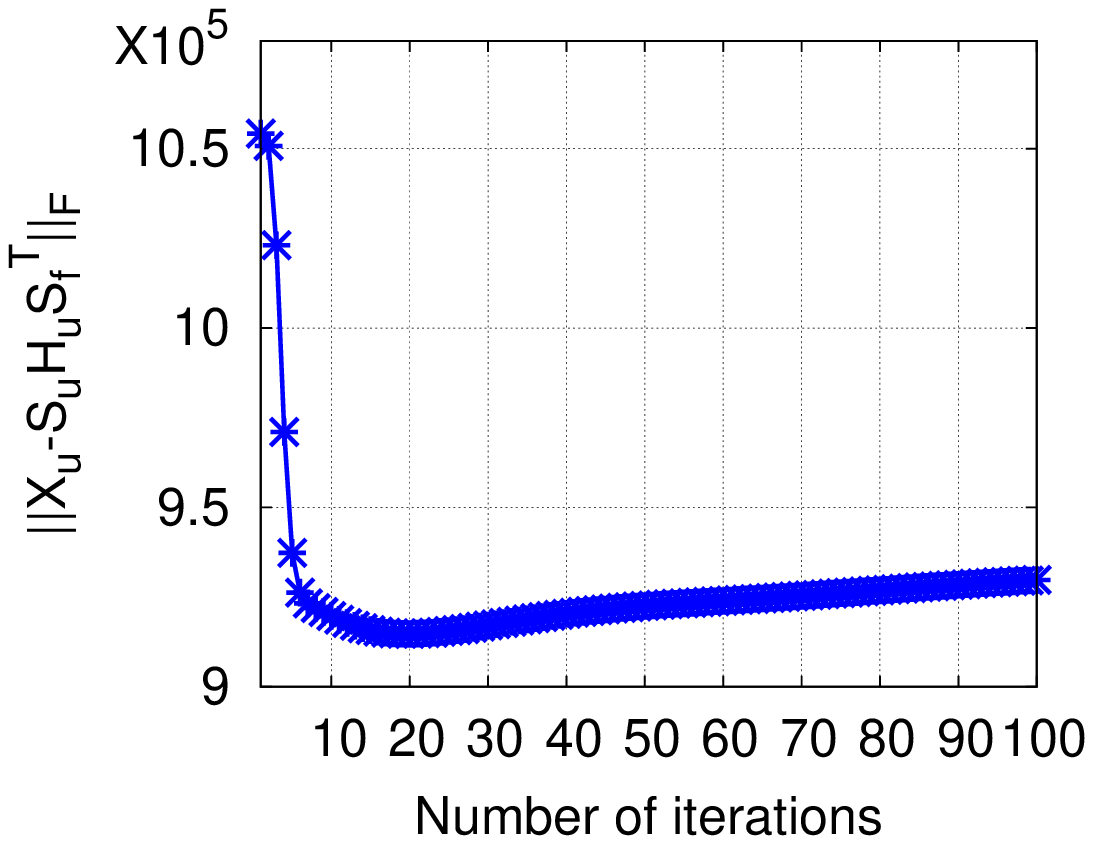}}
\subfigure[Average loss for Eq.~(\ref{equ:objective})]{\includegraphics[width=0.32\columnwidth]{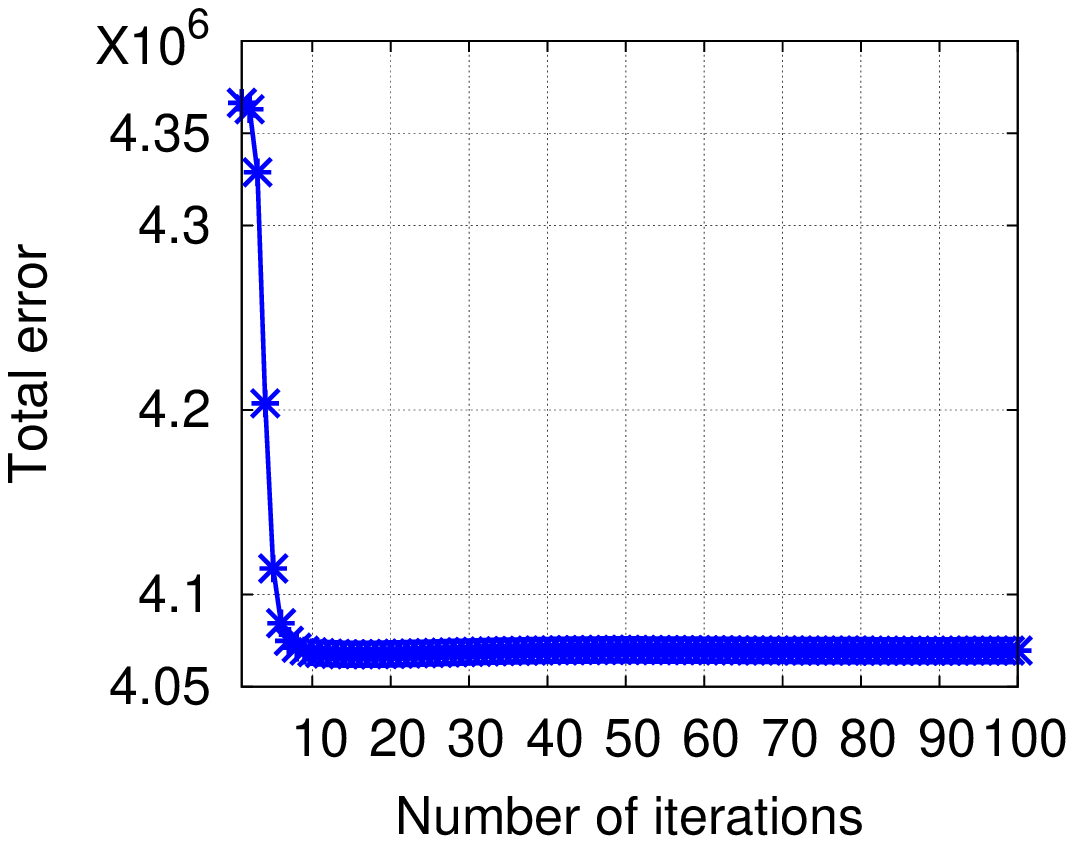}}
\caption{The average Frobenius loss of tweet-feature matrix approximation, user-feature matrix approximation and the objective function of tri-clustering on the Proposition 30 datasets}\label{fig:converge}
\end{figure*}

\noindent \textbf{Varying Parameters.} We first examine the effects of varying parameters on the performance of our approach. In the offline framework, we have two parameters, $\alpha$ and $\beta$, which control the contributions of feature lexicon information and the social relationship information, respectively. Since it is not clear whether these two parameters are independent or not, we vary these two parameters together, and then observe the change of both tweet-level and user-level performance.

Figure~\ref{fig:user-para} compares the user-level performance of different parameters over the Proposition 30 dataset. First, we observe that the best accuracy (colored as dark blue in Figure~\ref{fig:user-para}) can be found within either $\alpha$=0, $\beta=[0.5, 0.8]$,  or $\alpha=[0.7,1]$, $\beta=1$; while the best NMI is achieved with parameters $\alpha$=0, $\beta=[0.6, 0.9]$. Thus, the best parameters for user-level sentiment analysis can be set to $\alpha$=0, $\beta=[0.5, 0.8]$. The combination indicates that lexicon-based regularization is inessential for user-level sentiment analysis, especially when it is compared to social relation-based graph regularization. Despite graph regularization is useful (e.g., when $\beta=[0.5, 0.8]$), heavy regularization (i.e., $\beta$=1) deteriorates the performance. The results verify our assumption that social relation is useful, but they cannot replace the roles of latent feature vectors (e.g, tf-idf term vector representation) and user-tweet relations.

Now let us turn our attention to the relation between tweet-level quality and parameters. We want to answer the following two questions: whether the best combination of parameters ($\alpha$=0.0, $\beta$=[0.5, 0.8]) still holds, and whether the two regularizations are also needed for tweet-level analysis. Figure~\ref{fig:tweet-para} depicts the tweet-level sentiment results. When $\alpha$=0.1, $\beta$=0.9, tri-clustering obtains the best accuracy for Proposition 30; while $\alpha$=0.1, $\beta$=0.8, it achieves the best performance in terms of NMI. Thus, we suggest that a combination of parameters ($\alpha$=0.1, $\beta$=[0.8, 0.9]) might be a good choice. This differs from user-level sentiment analysis in that it prefers light lexicon-based regularization to no regularization ($\alpha$=0). Another difference is that tweet-level sentiment analysis is less sensitive to parameters than user-level sentiment analysis. When varying parameters, the tweet-level accuracy only varies between 81\% and 82\%, while the user-level accuracy changes from 73\% and 85\%.

We also conducted this set of experiments with varying parameters on another topic Proposition 37 dataset, and we observed similar results and hence omit the details here due to space limit. Moreover, in order to balance between the tweet-level performance and user-level performance, in all the following offline experiments, we set $\alpha$=0.05, $\beta$=0.8.

\vspace{0.2cm}\noindent\textbf{Convergence Analysis.} Figure~\ref{fig:converge} examines how fast our algorithm converges. When the number of iterations is around 10, our algorithm tends to converge in terms of total error of Eq.~(\ref{equ:objective}). However, if we look at each single component, after 10 iterations, first the algorithm minimizes the loss for Eq.~(\ref{equ:user}) at the cost of increasing the error of Eq.~(\ref{equ:tweet}), and then vice versa. This is because our objective is to minimize all components (including regularization and co-factorization), instead of each single equation. Thus, the algorithm searches among each local optimum of the five components and finally finds the global balancing point.

\vspace{0.2cm}\noindent\textbf{Comparison with Existing Methods.} Table~\ref{tab:baselinetweet} reports the tweet-level accuracy and NMI values of different methods. Although our tri-clustering method is worse than the supervised methods SVM and NB, it does not require any labeling, while the performance of a supervised method is highly related to the quality and sufficiency of labeled data. Compared with semi-supervised methods, our method is much better than LP with 5\% labels (i.e., LP-5), and better than or comparable with LP with 10\% labels (i.e., LP-10). Considering another state-of-the-art semi-supervised method UserReg, tri-clustering is worse than UserReg with 10\% labels (i.e., UserReg-10) on dataset Prop 30 but better than UserReg-10 on dataset Prop 37. These results support our claim that tri-clustering has an advantage over supervised and semi-supervised methods when labeled data are either insufficient or with poor quality. In another unsupervised method ESSA, two virtual tweet-tweet graph and feature-feature graph are built to improve the tweet-level accuracy. Specifically, two tweets or features are linked if they are similar to each other. The computation of tweet-tweet graph and feature-feature graph is very time consuming. Comparing our method with ESSA, our method focuses more on user-level accuracy and does not require the tweet-tweet graph and feature-feature graph, but the results show that our method is consistently better than ESSA in terms of both clustering accuracy and NMI values.

\begin{table}
  \centering
  \caption{Tweet-level sentiment analysis comparison}\label{tab:baselinetweet}
  \begin{tabular}{|l|c|c|c|c|}
  \hline
Metric&\multicolumn{2}{|c|}{Accuracy}&\multicolumn{2}{|c|}{NMI}\\
\hline
Prop&30 &37&30&37\\
\hline
\multicolumn{5}{|c|}{Supervised}\\
\hline
SVM~\cite{Topics2013}&89.35&93.17&--&--\\
NB~\cite{Go2009}&85.75&89.22&&\\
\hline
 \multicolumn{5}{|c|}{Semi-supervised}\\
 \hline
LP-5~\cite{Goldberg2006,Speriosu:2011}&77.20&87.49&--&--\\
LP-10~\cite{Goldberg2006,Speriosu:2011}&86.60&88.20&--&--\\
UserReg-10~\cite{confsdmDengHJLLW13}&86.76&90.08&--&--\\
\hline
\multicolumn{5}{|c|}{Unsupervised}\\
\hline
ESSA~\cite{WWW2013Hu}&81.69&85.87&38.71&15.88\\
Tri-clustering&81.87&92.15&40.24&18.93\\
Online tri-clustering&91.88&92.24&67.73&29.85\\
    \hline
  \end{tabular}
\end{table}

We also report the user-level performance comparison in Table~\ref{tab:baselineuser}. The clustering accuracy of tri-clustering is very close to the supervised methods SVM and NB. Compared with the semi-supervised methods, our method is significantly better than LP-10 and even outperforms UserReg-10. The better performance over UserReg-10 also indicates that the estimation of users' sentiments by aggregating over tweet-level sentiments is biased. Finally, our method is also significantly better than another unsupervised clustering approach BACG which also incorporates both user-user relations and feature vector representation of users.

\begin{table}[!t]
  \centering
  \caption{User-level sentiment analysis comparison}\label{tab:baselineuser}
  \begin{tabular}{|l|c|c|c|c|}
    \hline
Metric&\multicolumn{2}{|c|}{Accuracy}&\multicolumn{2}{|c|}{NMI}\\
\hline
Prop&30 &37&30&37\\
\hline
\multicolumn{5}{|c|}{Supervised}\\
\hline
SVM~\cite{Topics2013}&89.81&87.84&--&--\\
NB~\cite{Go2009}&88.69&83.8&&\\
\hline
\multicolumn{5}{|c|}{Semi-supervised}\\
 \hline
LP-5~\cite{Tan2011}&31.77&82.05&--&--\\
LP-10~\cite{Tan2011}&77.45&84.25&--&--\\
UserReg-10~\cite{confsdmDengHJLLW13}&82.10&84.28&--&--\\
\hline
\multicolumn{5}{|c|}{Unsupervised}\\
\hline
BACG~\cite{BACG12}&75.37&70.51&33.70&53.70\\
Tri-clustering&86.88&86.17&52.47&71.98\\
Online tri-clustering&89.22&88.48&53.89&73.48\\
    \hline
  \end{tabular}
\end{table}

\subsection{Online Performance Evaluation}

\begin{figure}[!t]
\centering
\subfigure[User-level]{\includegraphics[width=0.545\columnwidth]{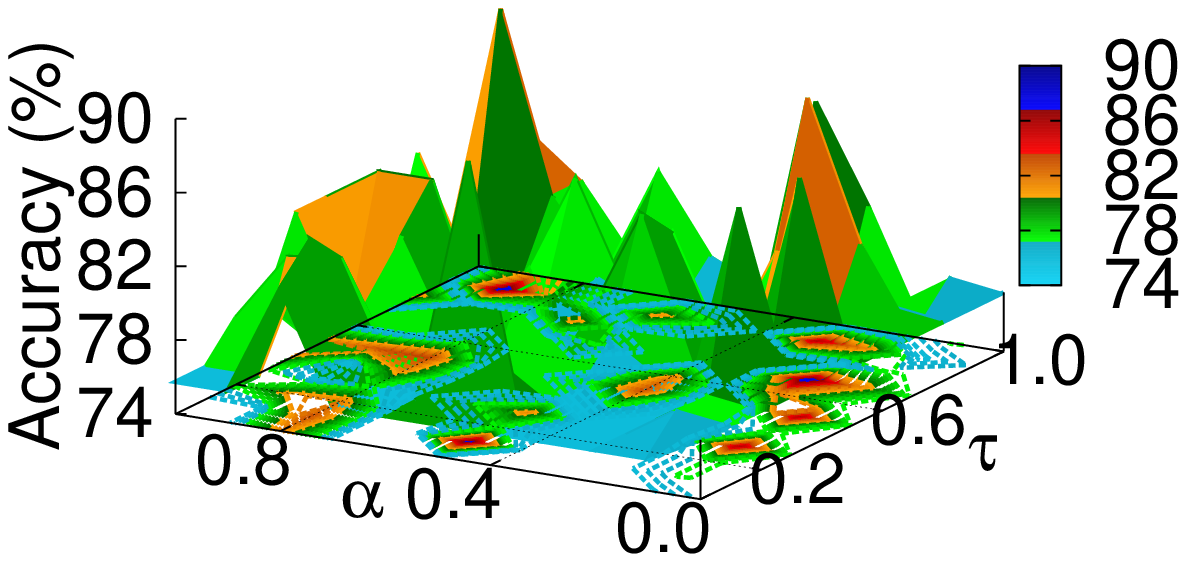}}
\hspace{-0.2cm}
\subfigure[Tweet-level]{\includegraphics[width=0.45\columnwidth]{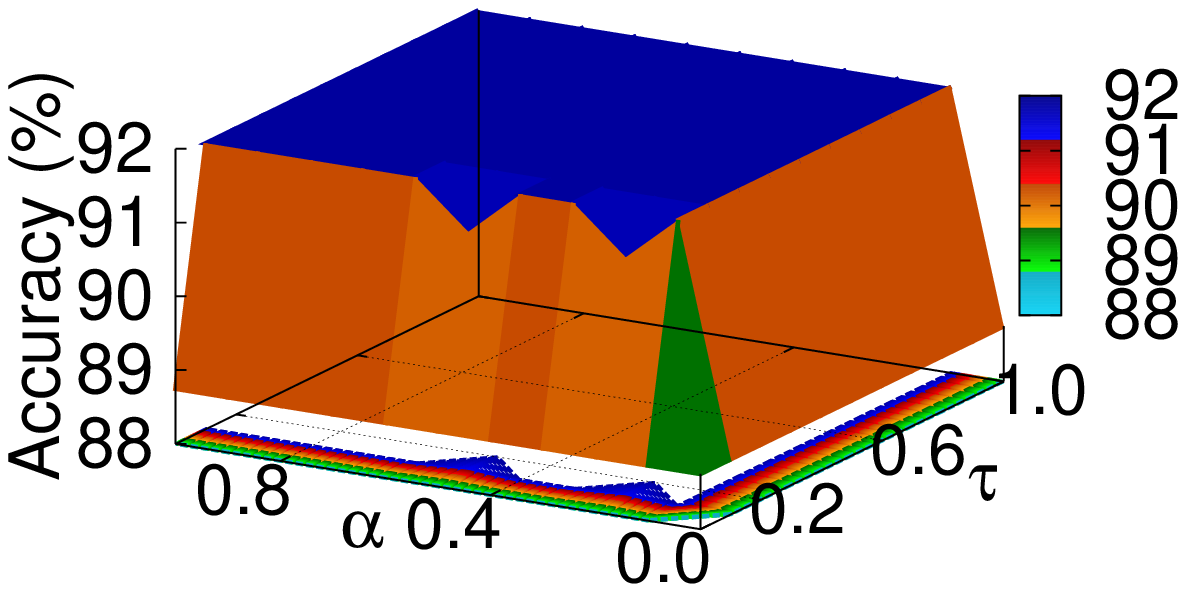}}
  \caption{Clustering accuracy when varying $\alpha$ and $\tau$ on Proposition 30 data (the figure is best viewed in color)}\label{fig:acc-tau}
\end{figure}

\begin{figure}[!t]
\centering
\includegraphics[width=0.65\columnwidth]{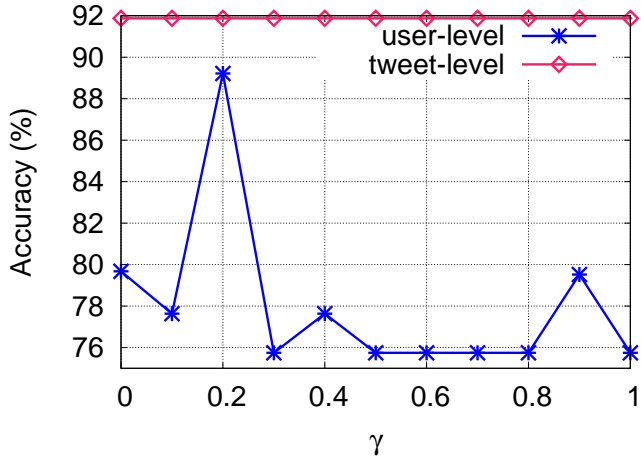}
  \caption{Clustering accuracy when varying $\gamma$ for Proposition 30 data}\label{fig:acc-gamma}
\end{figure}

\noindent \textbf{Varying Parameters.} We first evaluate the effects of varying the different parameters $\alpha$, $\beta$, $\gamma$, $\tau$, and time window size $w$, on the online performance of our algorithm. We notice that time window size $w$ is related to the granularity of timestamp. For example, using each second as the unit of timestamp naturally leads to larger $w$ than using each day as the unit. Consider the real scenario, in our experiments, we set the unit of timestamp as per day, and set $w=2$. We still set $\beta$=0.8, the same as the offline experiments but we need to change $\alpha$ since now it is related to parameter $\tau$ as well. Therefore, we first evaluate the performance when varying both $\alpha$ and $\tau$. The user-level accuracy and tweet-level accuracy are reported in Figure~\ref{fig:acc-tau}. In terms of user-level accuracy, it obtains the highest value when both $\alpha$ and $\tau$ are set to 0.9. In terms of tweet-level accuracy, similar to the offine experiments, it is much less sensitive unless both $\alpha$ and $\tau$ are greater than zero. We do not report the results of NMI values since we can draw similar conclusion as that from the accuracy values. Thus, we set $\alpha$ and $\tau$ as 0.9 for the online experiments.

The last step is to evaluate the effect of parameter $\gamma$ when all of the other parameters are fixed. The results are shown in Figure~\ref{fig:acc-gamma}. Clearly, the best result is obtained when $\gamma$=0.2. Meanwhile, parameter $\gamma$ does not have any effect on the tweet-level accuracy. This is because $\gamma$ controls the smooth evolution of user-level sentiment, which is relatively independent with the tweet-level sentiment.

\begin{figure*}[!t]
\centering
\includegraphics[width=\columnwidth]{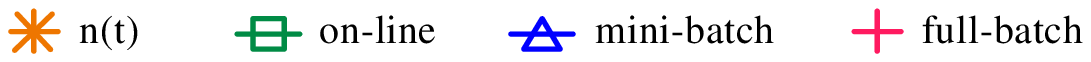}\\
\subfigure[Total running time]{\label{subfig:onlinetime30}\includegraphics[width=0.32\columnwidth]{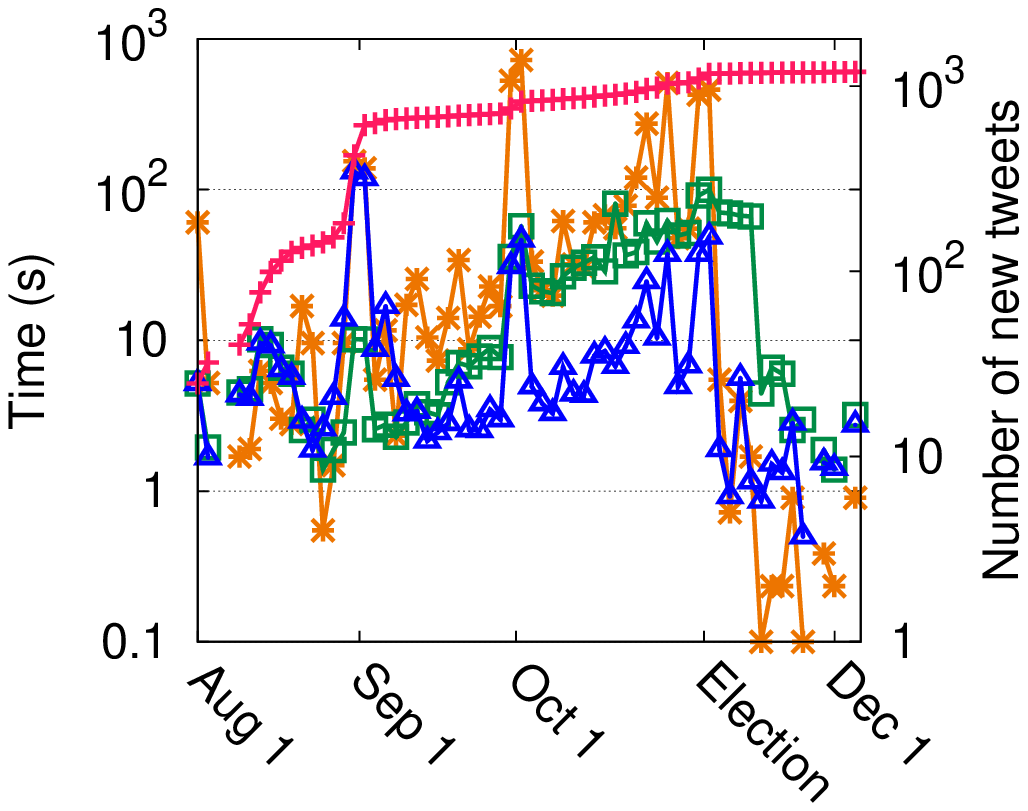}}
\hspace{-0.1mm}
\subfigure[Tweet-level accuracy]{\label{subfig:onlinetweet30}\includegraphics[width=0.32\columnwidth]{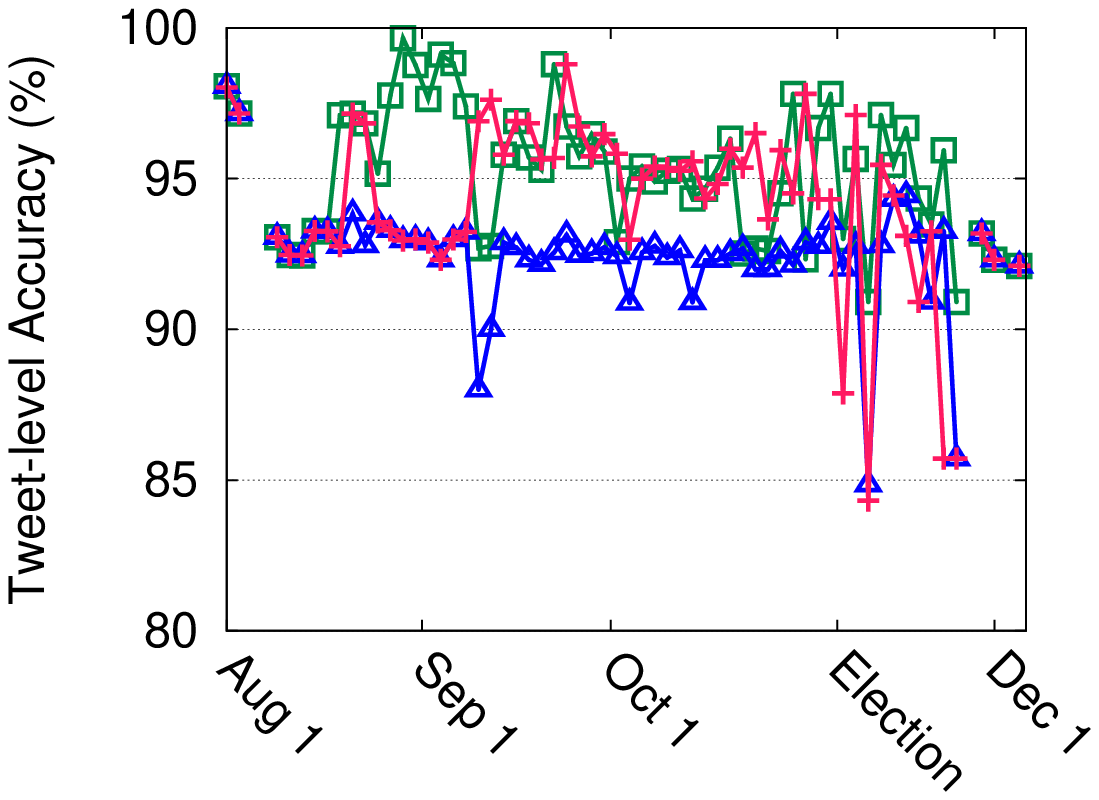}}
\hspace{-0.1mm}
\subfigure[User-level accuracy]{\label{subfig:onlineuser30}\includegraphics[width=0.32\columnwidth]{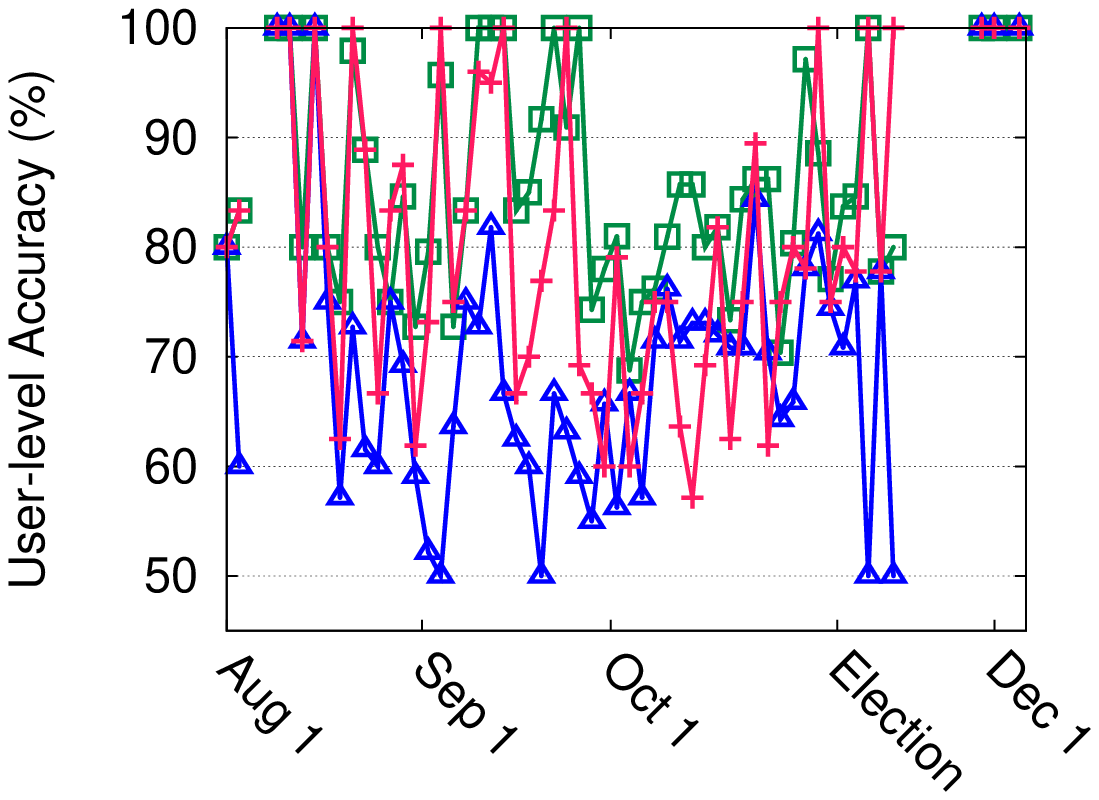}}
\caption{Online performance results for Proposition 30 data (the figure is best viewed in color)}\label{fig:onlinecompare30}
\end{figure*}

The results on another topic Proposition 37 dataset are similar and hence omit the details here due to space limit.

\vspace{0.2cm}\noindent\textbf{Comparison with the Offline Algorithm.} The last two rows of both Table~\ref{tab:baselinetweet} and Table~\ref{tab:baselineuser} show that the online algorithm obtains much higher accuracy than the offline algorithm for both tweet-level and user-level performance, and sometimes even outperforms the supervised methods. This is not surprising since in the online framework, we have considered the evolution of latent feature vectors while in the offline framework, it simply assumes that the features are static. However, in real life, people may tend to use different vocabularies in different time period. Thus, our online algorithm, which subsumes the evolution of latent feature vectors, outperforms the offline algorithm in terms of clustering accuracy.

%

\begin{figure*}[!t]
\centering
\includegraphics[width=\columnwidth]{figure/legend2}\\
\subfigure[Total running time]{\label{subfig:onlinetime37}\includegraphics[width=0.32\columnwidth]{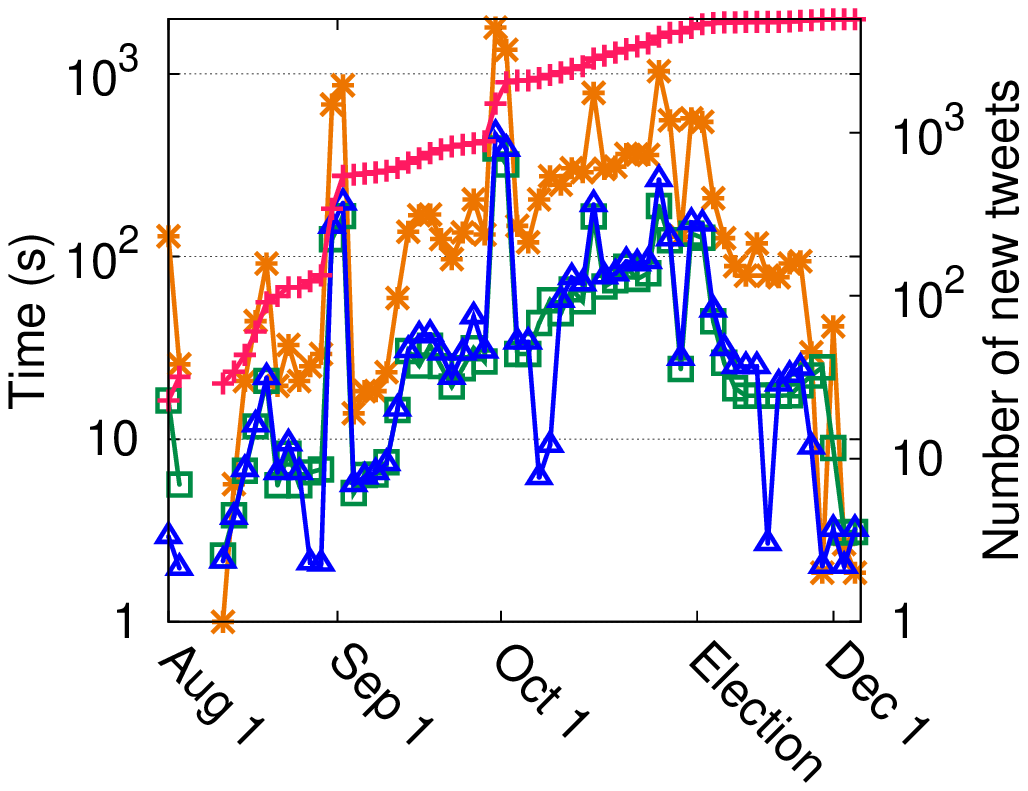}}
\hspace{-0.1mm}
\subfigure[Tweet-level accuracy]{\label{subfig:onlinetweet37}\includegraphics[width=0.32\columnwidth]{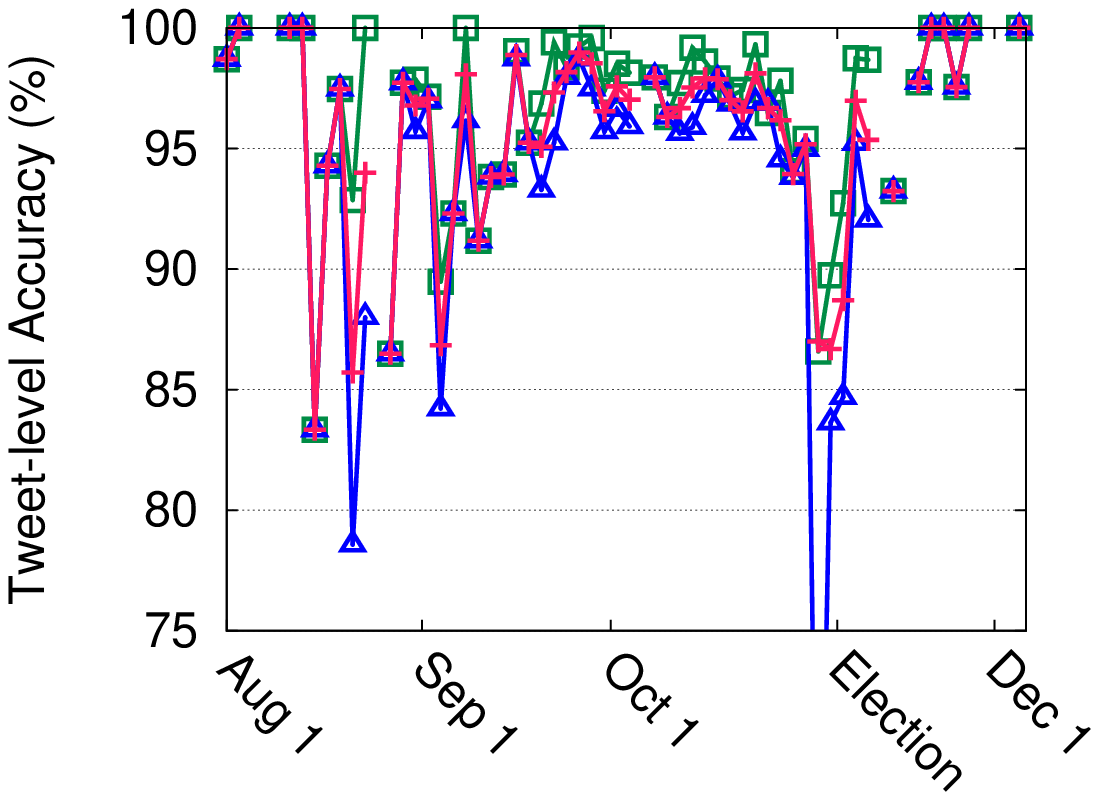}}
\hspace{-0.1mm}
\subfigure[User-level accuracy]{\label{subfig:onlineuser37}\includegraphics[width=0.32\columnwidth]{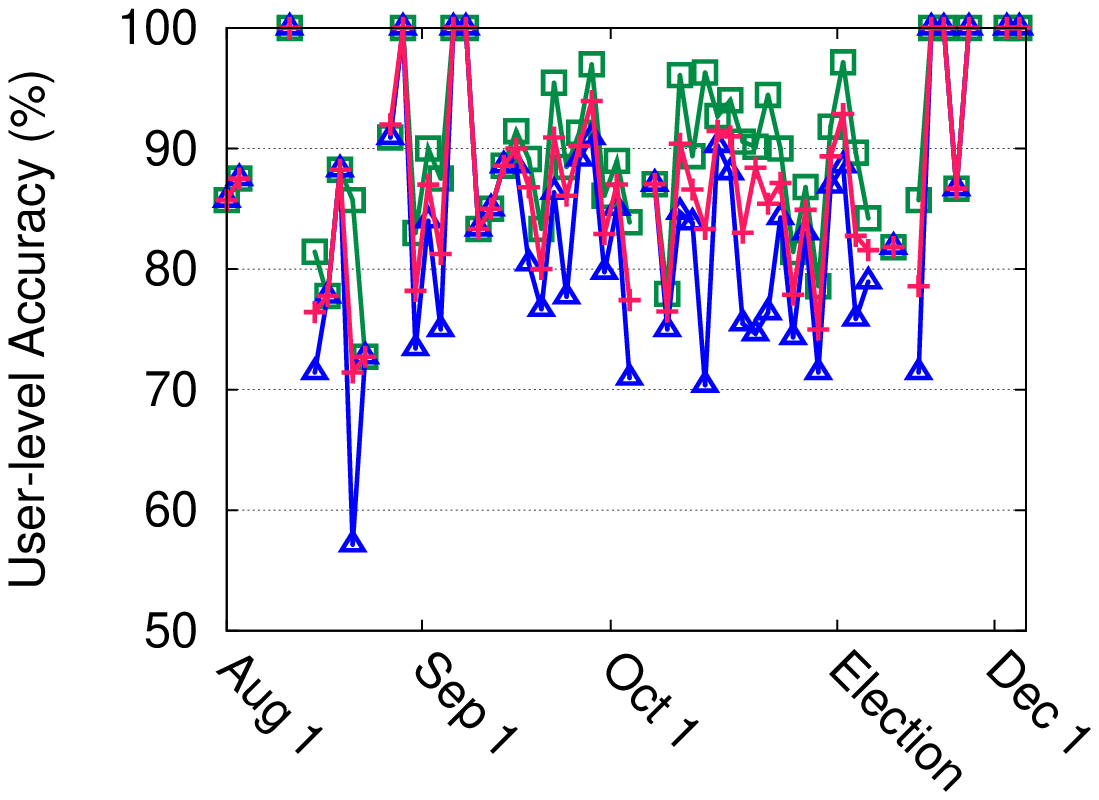}}
\caption{Online performance results for Proposition 37 data (the figure is best viewed in color)}\label{fig:onlinecompare37}
\end{figure*}

\vspace{0.2cm}\noindent\textbf{Performance Comparison of Online Algorithms.} We also compare our online algorithm with the two baseline algorithms, mini-batch and full-batch, on both the Proposition 30 and Proposition 37 datasets. The results are reported in Figures~\ref{fig:onlinecompare30} and~\ref{fig:onlinecompare37}. First, we verify whether our online algorithm is as efficient as the theoretical time complexity analysis. Figure~\ref{subfig:onlinetime30} and Figure~\ref{subfig:onlinetime37} plot the running time of the three algorithms. Note that the left y-axis represents the running time, and the right y-axis denotes the number of new tweets, $n(t)$, for each timestamp $t$. For Proposition 30, our online algorithm is much faster than the full-batch algorithm but slower than the mini-batch algorithm. This is because our online algorithm needs additional cost to access the previous clustering results (i.e., $S_{fw}(t)$, $S_u(t-1)$) and compute the temporal regularization. However, when they are tested on Proposition 37 data, the online algorithm is comparable to the mini-batch algorithm. The difference in performance for the different datasets is caused by the variation on the average number of new tweets per day (i.e., $n(t)$) between Proposition 30 and Proposition 37. Based on Figure \ref{subfig:onlinetime30} and Figure~\ref{subfig:onlinetime37}, we notice that the average number of new tweets per day on Proposition 37 is much larger than Proposition 30; and thus the cost to access the data matrices $X_p(t)$, $X_u(t)$ and $X_r(t)$ dominate the incurred overhead by the matrices $S_{fw}(t)$ and $S_u(t-1)$. When $n(t)$ is small such as on Proposition 30, though our online algorithm is slower than the mini-batch algorithm, it is less sensitive to the bursty of tweets than the mini-batch algorithm. For example, around Sep 1, there is a sudden rise in the volume of tweets and our online algorithm is much more efficient than the mini-batch algorithm due to its advantage in utilizing the previous clustering results.

Figures~\ref{subfig:onlinetweet30} and~\ref{subfig:onlinetweet37} compare the tweet-level accuracy of the three algorithms. The mini-batch algorithm, which is very efficient since it only needs to access a small set of newly arrived data at each timestamp, obtains the lowest accuracy. Our online algorithm, achieves an accuracy as good as the expensive full-batch algorithm on proposition 30 data and even outperforms the full-batch algorithm on proposition 37.

The user-level performance comparison on Proposition 30 and 37 is reported in Figures~\ref{subfig:onlineuser30} and~\ref{subfig:onlineuser37}. The results are similar to the tweet-level accuracy, which show that our online algorithm is comparable to the full-batch algorithm, and both of them are significantly better than the mini-batch algorithm.

To conclude, the results verify that our online algorithm achieves a good trade-off between efficiency and clustering quality.

\section{Related Works}\label{sec:related}
We first discuss related work on sentiment analysis. We also discuss some related work on non-negative matrix factorization.
\subsection{Sentiment Analysis}
We summarize a set of representative (but by no means exhaustive) methods  to sentiment analysis in Table~\ref{tab:related}, where we group existing approaches into three directions. First, we consider whether a method aims to identify positive or negative sentiment in a piece of text (tweet-level analysis) or to determine the sentiments of users (user-level analysis). A large amount of research in the area of sentiment analysis has focused on classifying text polarity~\cite{MelvilleKDD2009,Barbosa2010,Go2009,Davidov2010,Wangcikm2011,HatzivassiloglouACL97,Speriosu:2011,NguyenWSDM2012,WWW2013Hu,ZhuGPLDS13}. Smith et al.~\cite{Topics2013} and Deng et al.~\cite{confsdmDengHJLLW13} analyzed the sentiments of users by aggregating the sentiments of their tweets. Tan et al.~\cite{Tan2011} directly analyzed the sentiments of users using a semi-supervised approach. Specifically, a semi-supervised label propagation algorithm is utilized to determine the sentiment of a user by the sentiments of his/her tweets and the sentiments of his/her immediate neighbor users in a heterogeneous graph built upon social relations. However, with insufficient labeled nodes or the labeled nodes are densely condensed in a small region of the entire graph, the performance of this approach is not encouraging. Another issue is that Smith et al.~\cite{Topics2013} have pointed out that the emotion correlation among users and following or @mention users (which are used in~\cite{Tan2011} to build heterogeneous graph), is relatively lower than users and re-tweeting users. Kim et al.~\cite{icwsm2013} utilized the collaborative filtering techniques to analyze the sentiments of users based on the sentiments of similar users. The similarity of two users are evaluated by whether they have expressed similar sentiments towards the same set of topics. This approach totally ignored the rich information of tweets and features, as well as social relationship such as user-user re-tweeting relation. Instead, in this work, we propose a tri-clustering framework, to obtain the sentiment clustering of both tweets and users simultaneously. Our approach utilizes the re-tweeting social relation and dependencies among users, tweets, and features, and is independent with the quality of labeled data.

\begin{table}
\centering
\caption{Methods for sentiment analysis }\label{tab:related}
 \begin{tabular}{|l|l|l|l|l|l|l|l|l|}
\hline
\small{methods}&\small{\cite{Barbosa2010,Davidov2010}}&\cite{Goldberg2006}
&\small{\cite{Topics2013}}&\small{\cite{icwsm2013}}&\small{\cite{confsdmDengHJLLW13}}&\small{\cite{Castellanos2011}}&\small{\cite{HatzivassiloglouACL97}}&\small{this}\\
&\small{\cite{Go2009,PangEMNLP02}}&\cite{Speriosu:2011}&&\small{\cite{Tan2011}}&&\small{\cite{NguyenWSDM2012}}&\small{\cite{WWW2013Hu}}&\small{work}\\
&\cite{Wangcikm2011,Lin2012}&&&&&&&\\
\hline
\small{tweet}&$\surd$&$\surd$&$\surd$&&$\surd$&$\surd$&$\surd$&$\surd$\\
\hline
\small{user}&&&$\surd$&$\surd$&$\surd$&&&$\surd$\\
\hline
\small{SL}&$\surd$&&$\surd$&&&&&\\
\hline
\small{USL}&&&&&&$\surd$&$\surd$&$\surd$\\
\hline
\small{SSL}&&$\surd$&&$\surd$&$\surd$&&&\\
\hline
\small{dynamic}&&&&&&$\surd$&&$\surd$\\
\hline
\end{tabular}
\vspace{-0.3cm}
\end{table}

Second, we focus on the requirement of labeled data by different methods. Many existing approaches~\cite{Barbosa2010,Go2009,Davidov2010,Wangcikm2011,MelvilleKDD2009,Topics2013} are based on supervised learning (SL), where, give some labeled data, one trains and applies a standard classifiers to extract sentiment. Graph-based semi-supervised learning (SSL) algorithms, have been applied to the problem of sentiment analysis~\cite{Goldberg2006,Speriosu:2011,Tan2011,confsdmDengHJLLW13}, assuming that similar texts/users receive similar sentiment labels.  Unfortunately, the above approaches either use some forms of linguistic processing~\cite{Barbosa2010,Davidov2010} or rely on large amounts of training data~\cite{Go2009,Lin2012,Topics2013}, both of which heavily involve the inputs from humans\footnote{Human labeling is very difficult even with the help of Crowdsourcing.}. In contrast to previous approaches, unsupervised learning (USL) approaches, which do not require any labeling or input from human, are inadequately studied. Some approaches\cite{HatzivassiloglouACL97, NguyenWSDM2012} used linguistic models or some other form of knowledge to categorize the sentiment of documents. A more recent approach ESSA~\cite{WWW2013Hu} studied  the problem of unsupervised
sentiment analysis with emotional signals. In this work, we establish the duality between sentiment clustering and co-clustering of a tripartite graph with the unified tri-clustering framework, which belongs to the category of unsupervised methods.

Finally, we focus on the ability of different methods to perform dynamic sentiment analysis. Most of the existing approaches discussed above either just perform static sentiment analysis or simply report how the number of positive/negative tweets are changing over time~\cite{Castellanos2011,NguyenWSDM2012}. Instead, our framework is able to identify the evolution of both features and sentiments of users, which distinguishes our work from existing sentiment analysis works.

\subsection{Non-negative Matrix Factorization}
Due to its wide application in various areas such as text mining~\cite{ZhuangSADM2011,WWW2013Hu}, pattern recognition~\cite{CaiPAMI2011}, machine learning~\cite{GuKDD2009,LongWDSYAAAI12} and bioinformatics~\cite{devarajan2008non}, nonnegative matrix factorization (NMF) has attracted much interest from researchers. Generally, nonnegative matrix factorization aims to factor a matrix $X$, into two~\cite{LeeNIPS2000,CaiPAMI2011} (or three~\cite{DingKDD2006,ZhuangSADM2011,WWW2013Hu}) lower-dimension matrices and minimizes the square error/divergence between $X$ and the approximation of $X$ using those lower-dimension matrices. There are several algorithms that are proposed to find the sub-optimal solution of those lower-dimension matrices, for instance, Lee and Seung~\cite{LeeNIPS2000} proposed two different multiplicative algorithms to update the matrices. Other more recent approaches include using the projected gradient descent methods~\cite{LinNC2007}, the active-set method and the block principal pivoting~\cite{KimSIAMJSC2011} to update the matrices. If one of the factors (lower dimension matrices) satisfies the separability condition, Arora et al.~\cite{AroraSTOC2012} also proposed a polynomial-time algorithm to find the exact NMF solution.

In this work, we explored the feasibility of applying NMF to Twitter sentiment analysis domain. We further developed an online framework with several temporal and graph regularization for both user-level and tweet-level sentiment analysis. Both our online objective function and optimization algorithm are different from the existing online NMF algorithms~\cite{CaoIJCAI2007,WangLK11,SahaWSDM2012}.

\section{Conclusions}\label{sec:conclude}

We studied both user-level and tweet-level dynamic sentiment analysis on social media data. We proposed a novel tri-clustering framework, making use of mutual dependency among features, tweets and users. We developed an analytical algorithm with fast convergence to solve the proposed objective function in the offline tri-clustering framework. We further investigated how the proposed framework can be extended to online setting by considering the dynamic evolution of features and sentiments of users. We proposed an online algorithm, which is efficient in terms of both running time and storage size, to achieve a good clustering for both tweets and users in a dynamic setting. We also extensively evaluated our algorithms on real Twitter data about November 2012 California ballot initiatives.

For future work, we consider to propose a unified tripartite graph co-clustering framework, with a set of optional regularizations which include graph regularization, sparsity regularization, diversity regularization, temporal regularization, and guided regularization (semi-supervised regularization). This framework can be applied to many different domains such as community detection, transfer learning and node role mining, without the restriction to only sentiment analysis.
\bibliographystyle{abbrv}
\bibliography{OpinionMining}
\end{document}